\documentclass[a4paper,10pt,notitlepage,twocolumn]{revtex4-1}
\usepackage[utf8]{inputenc}
\usepackage[ngerman,english]{babel}
\pdfoutput=1
\usepackage{url}
\usepackage{bbm}
\usepackage[T1]{fontenc}
\usepackage[utf8]{inputenc}
\usepackage{graphicx}
\usepackage{eurosym}
\usepackage{longtable}
\usepackage{amsmath}
\usepackage{amssymb}
\usepackage{hyperref}
\usepackage{amsthm}
\usepackage{subfigure}
\usepackage{setspace}
\usepackage{colortbl}
\usepackage[titletoc,title]{appendix}
\usepackage[left=2.0cm,right=2.0cm,top=3.0cm,bottom=3.0cm]{geometry}
\begin{document}
 \title{A hybrid quantum repeater for qudits}
\author{Marcel Bergmann}
\author{Peter van Loock}
\affiliation{Institut für Physik, Johannes Gutenberg-Universität, Staudingerweg 7, 55128 Mainz, Germany}
\begin{abstract}
\noindent
We present a "hybrid quantum repeater" protocol for the long-distance distribution of atomic entangled states beyond qubits. In our scheme,
imperfect noisy entangled pairs of two qudits, i.e., two discrete-variable $d$-level systems, each of, in principle, arbitrary dimension
$d$, are initially shared between the intermediate stations of the channel. This is achieved via local, sufficiently strong light-matter interactions,
involving optical coherent states and their transmission after these interactions, and optical measurements on the transmitted field modes, especially 
(but not restricted to) efficient continuous-variable homodyne detections ("hybrid" here refers to the simultaneous exploitation
of discrete and continuous variable degrees of freedom for the local processing and storage of entangled states
as well as their non-local distribution, respectively). For qutrits we quantify the light-matter entanglement that can be effectively
shared through an elementary lossy channel, and for a repeater spacing of up to 10 km we show that the realistic (lossy) qutrit entanglement
is even larger than any ideal (loss-free) qubit entanglement.
After including qudit entanglement purification and swapping procedures,
we calculate the long-distance entangled-pair distribution rates and the final entangled-state fidelities for total
communication distances of up to 1280 km. With three rounds of purification, entangled qudit pairs of near-unit fidelity can be distributed
over 1280 km at rates of the order of, in principle, 100 Hz.
\end{abstract}
\maketitle
\section{Introduction}
\label{sec: intro}
Long-distance quantum communication is one of the most challenging tasks in practical quantum information. For future quantum networks, 
the distribution of entanglement between widely separated parties is necessary to make teleportation and secure communication over long distances
possible. In practice, however, the direct transmission of quantum information or entangled states is performed by sending light through a lossy quantum channel, which leads to 
an exponential decay of the success rate or the fidelity. To overcome this problem, quantum repeaters were proposed \cite{Briegel1, Briegel2, Sangouard}. 

From the perspective of the most recent quantum repeater research, 
a quantum repeater protocol can be classified into three distinct categories, referred to as quantum repeater generations \cite{Optiarchi, Ultrafast}. Though
much slower compared to second and third generation quantum repeaters based on quantum error correction of, respectively,
local (operation and memory) or, in addition, transmission errors, first generation quantum repeaters are 
attractive due to their immediate experimental feasibility (however, for a fairly practical approach to a third generation quantum repeater,
see \cite{Ewert1,Ewert2}). In first generation quantum repeaters, by means of entanglement swapping \cite{swap}, the distribution of long-distance entanglement
is achieved via initial short-distance entanglement distributions. Hence, for the 
realization of first generation quantum repeater schemes, the heralded generation of short-distance entanglement and the availability of quantum
memories are essential prerequisites.\\
A prominent instance of a first generation quantum repeater scheme is the well-known DLCZ protocol \cite{DLCZ}
which uses atomic ensembles as quantum memories and single photons with linear optics for entanglement distribution and swapping.
A remarkable feature of the DLCZ scheme is that the so-called purification of entanglement, 
turning imperfect mixed entangled states into purer (in principle, perfect) versions of entangled states,
is built into the process of entanglement distribution and swapping (purifying the entangled atomic ensembles from the effects of transmission
and memory losses, respectively). Otherwise, in a standard first generation quantum repeater \cite{Briegel1, Briegel2}, quantum
error detection must be included via additional rounds of entanglement purification acting on two
or more copies of entangled states and employing local quantum logic (together with two-way classical communication).
Second generation schemes use quantum error correction against memory errors, while in third generation quantum repeaters
no memories are necessary \cite{Munro}, since, for example, suitably encoded quantum information is directly sent through the channel \cite{Ultrafast, Optiarchi,Ewert1,Ewert2}. 
A conceptually distinct version of such a loss-error-correction-based repeater is the all-optical
scheme of Azuma et al. \cite{Azuma} based on the distribution of entangled cluster states. This scheme also relies on sufficiently fast feedforward operations
(as opposed to the all-optical scheme of \cite{Ewert1, Ewert2}).\\
All experimental demonstrations to date are for elements of a first generation repeater,
although light-matter interfaces and/or memories are still too inefficient 
to exceed the bounds \cite{Takeoka, Pirandola} of repeaterless quantum communication (or to even  scale up a repeater to really large distances).
In fact, almost all quantum memories that have been demonstrated so far perform worse compared to a simple optical fiber loop \cite{Cho}.
\\
A suitable first generation "hybrid quantum repeater" (HQR) protocol for the distribution of atomic qubit-qubit entanglement was given in \cite{HQR, Bright, Ladd}.
Similar to other hybrid quantum information processing schemes \cite{HybridQIP}, this protocol combines the advantages of  discrete and continuous 
variable quantum states. Atomic two-level systems with long coherence time serve as quantum memories while optical coherent states are used to 
generate the initial entanglement between the atoms using dispersive light-matter interactions and, in particular,  highly efficient homodyne measurements. 
Employing such Gaussian measurements and Gaussian states as the initial resources appears very attractive
from a practical point of view compared to repeater schemes based on the generation and detection of single photons.
A particular  experimental approach to this scheme, based on ions, was 
considered in \cite{Pfister}. Another, similar HQR protocol can be found in \cite{Brask} and a recent hybrid approach to entanglement 
swapping using coherent states and linear optics is presented in \cite{Jeongswap}.\\
On the fundamental level, higher-dimensional quantum systems of dimension $d$, so called qudits, do not only play an important role
in closing of the detection loophole in Bell test experiments \cite{violation1,violation2}. In addition,
it has been shown in \cite{Cerf} that qudits lead to an increase in data transfer and especially to a higher security 
in quantum key distribution (QKD) \cite{Scarani} compared to schemes involving only qubits \cite{Twist}. One possibility to realize
such improved schemes is the initial distribution of high-dimensional entanglement using correspondingly high-dimensional quantum repeaters,
which is the topic of this paper \footnote{Note
that inferring from the results of Refs. \cite{Takeoka, Pirandola}, e.g., the effective secret bit rate in a long-distance QKD scheme based on direct state transmissions
cannot be improved beyond that of, for instance, a qubit-based BB84 scheme. Thus, on a fundamental level, beyond-qubit-type encodings do not seem
to be particularly useful for direct long-distance QKD applications. Nonetheless, when employing quantum repeaters, switching to qudits may indeed be useful.}. Despite the many existing works on qubit quantum repeaters, rather little attention has been paid
to qudit quantum repeaters aiming at the long-distance-distribution of qudit entanglement and information.\\
In this paper, we generalize the HQR protocol for the distribution of qubit-qubit entanglement \cite{HQR, Bright} to the case of
qudit-qudit entanglement, i.e., bipartite states of multilevel systems. 
The structure of the paper is as follows: in Sec. \ref{sec: qubit}, we review the HQR protocol for the qubit case
and adapt it to our later generalization for qudits. In Sec. \ref{sec: qutrit}, we generalize this
scheme to the case of three-level systems (qutrits). After proposing a generalized dispersive qutrit-light interaction, we discuss the process of
entanglement generation in elementary links using this interaction. We consider both homodyne detection and unambiguous state discrimination (USD) for the measurement on the light
mode. Including entanglement purification for the initial qutrit-qutrit entangled states, we calculate the final rates and fidelities for our generalized
entanglement distribution scheme in various scenarios.
Based on these results, in Sec. \ref{sec: qudit} we discuss a generalization to arbitrarily dimensional quantum systems before we conclude
in Sec. \ref{sec: conclusions}.
\section{Hybrid quantum repeater for qubits}
\label{sec: qubit}
\begin{figure}[t!]
\centering
\includegraphics[width=0.4\textwidth]{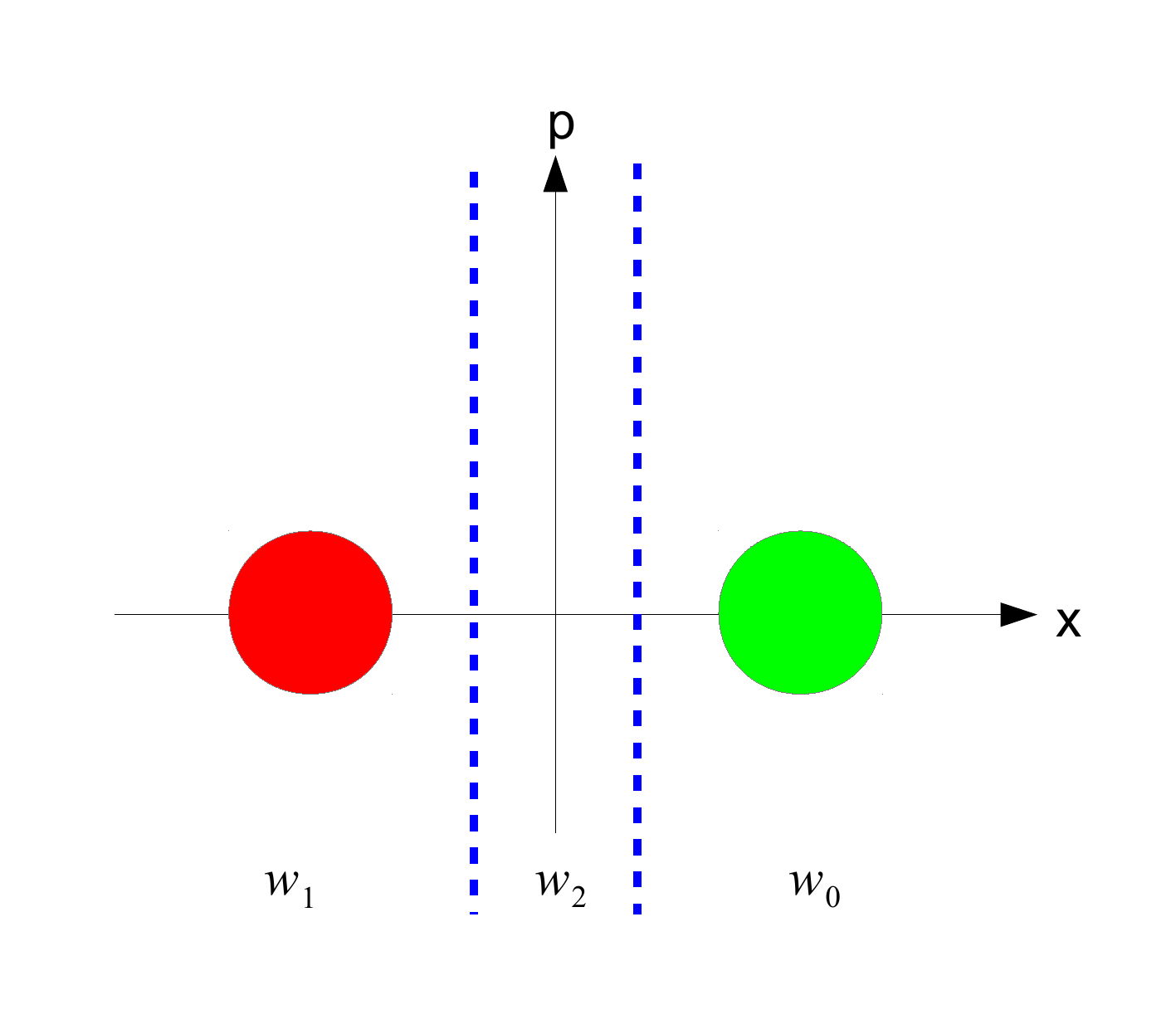}
\caption{Phase space representation of two coherent states $|\sqrt{\gamma}\alpha\rangle$ and $|-\sqrt{\gamma}\alpha\rangle$
to be distinguished by homodyne detection. The measurement window $w_{2}$ includes all failure events that are discarded.}
\label{fig: qubitwindows}
\end{figure}

The physical setup for a qubit HQR is as follows:
the qubit is represented by the two spin states $|0\rangle$ and $|1\rangle$ of an atomic electron. The atom is placed 
into a cavity and the electronic spin interacts with a bright coherent-state light pulse. The situation
at hand is theoretically described by the Jaynes-Cummings model in the limit of large detuning \cite{JC}, i.e.,
the probe pulse and the cavity are in resonance, but both are detuned from the resonance frequency of the electronic transition.\\
The interaction Hamiltonian in this model reads $H_{int}^{(2)}=\hbar g \sigma_{z} a^{\dagger}a$, where 
$\sigma_{z}=-\frac{1}{2}|0\rangle\langle0|+\frac{1}{2}|1\rangle\langle 1|$ corresponds to a Pauli operator
on the spin state and $a^{\dagger}a$ is the photon number operator of the light mode. Furthermore, the parameter $g$
describes the strength of the spin-light coupling.\\
Based on this interaction Hamiltonian, the corresponding unitary transformation is given by $U_{2}(\theta)=\exp(i\theta\sigma_{z}a^{\dagger}a)$
(with  an effective interaction time  $\theta=gt$) and, up to an unconditional phase shift of the mode by $e^{i\theta/2}$, acts on the spin-light system effectively as a 
controlled phase rotation, i.e.
\begin{equation}
\label{eq: qubitdis}
 U_{2}(\theta)[(|0\rangle+|1\rangle)\otimes|\alpha\rangle]=|0\rangle|\alpha\rangle+|1\rangle|\alpha e^{i\theta}\rangle.
\end{equation}
In the literature, this interaction is also known as dispersive interaction \cite{gerry}.
For the generalization that we are aiming at, we consider the case $\theta=\pi$, corresponding to a strong interaction resulting
in coherent states $|\pm \alpha\rangle$ on the light mode.\\
The repeater protocol now works as follows: the matter system is prepared in the state $|0\rangle+|1\rangle$ and interacts dispersively
with a single-mode coherent state $|\alpha\rangle$ (referred to as "qubus") as described by Eq. \eqref{eq: qubitdis}. Note that this 
leads to a pure (effectively qubit-qubit) entangled state between the light mode and the matter system.\\
The light mode is then sent through an optical channel where it inevitably suffers from photon loss. The photon loss can be modeled
by mixing the light mode with a vacuum state at a beam splitter with transmittance $\gamma$, where $1-\gamma$
is related to the loss probability of a single photon. It is also related to the optical propagation distance  $L$, i.e., $\gamma=\exp\left(-\frac{L}{L_{att}}\right)$
with the attenuation length $L_{att}\approx 22~\text{km}$ for photons at telecom wavelength. 
\\
After applying the beam splitter, the total pure state of the matter system, the qubus light mode and the loss mode reads as
\begin{equation}
 \frac{1}{\sqrt{2}}(|0\rangle|\sqrt{\gamma}\alpha\rangle|\sqrt{1-\gamma}\alpha\rangle+|1\rangle|-\sqrt{\gamma}\alpha\rangle|-\sqrt{1-\gamma}\alpha\rangle).
\end{equation}
\noindent
The relevant joint state of the matter system and the light mode is obtained by tracing out the loss mode. Since the coherent states
$|\alpha\rangle$ and $|-\alpha\rangle$ are not orthogonal, it is useful to transform these into an orthogonal basis. A suitable orthogonal basis
in this case is the basis of even and odd cat states (throughout we assume $\alpha\in \mathbb{R}$),
\begin{eqnarray} 
\label{eq: qubitbasis1}
 |u\rangle&=\frac{1}{\sqrt{N_{u}(\alpha)}}(|\alpha\rangle+|-\alpha\rangle),\\
 \label{eq: qubitbasis2}
 |v\rangle&=\frac{1}{\sqrt{N_{v}(\alpha)}}(|\alpha\rangle-|-\alpha\rangle),
\end{eqnarray}
with normalization constants $N_{u}(\alpha)=2(1+e^{-2\alpha^{2}})$ and $N_{v}(\alpha)=2(1-e^{-2\alpha^{2}})$. Expressed in this basis, one has
\begin{eqnarray}
 |\alpha\rangle=\frac{1}{2}(\sqrt{N_{u}(\alpha)}|u\rangle+\sqrt{N_{v}(\alpha)}|v\rangle),\\
 |-\alpha\rangle=\frac{1}{2}(\sqrt{N_{u}(\alpha)}|u\rangle-\sqrt{N_{v}(\alpha)}|v\rangle).
\end{eqnarray}
After tracing out the loss mode in this basis, the resulting state of the matter system and the qubus light mode becomes
\begin{equation}
\label{eq: qubitchannel}
\begin{aligned}
 \rho_{out}&=\frac{N_{u}(\sqrt{1-\gamma}\alpha)}{4}\\
 &\times\left[\frac{1}{\sqrt{2}}\left(|0\rangle|\sqrt{\gamma}\alpha\rangle+|1\rangle|-\sqrt{\gamma}\alpha\rangle\right)\right]\times H.c.\\
 &+\frac{N_{v}(\sqrt{1-\gamma}\alpha)}{4}\\
 &\times\left[\frac{1}{\sqrt{2}}\left(|0\rangle|\sqrt{\gamma}\alpha\rangle-|1\rangle|-\sqrt{\gamma}\alpha\rangle\right)\right]\times H.c.
\end{aligned}
 \end{equation}
This is a mixed entangled state between the matter system (the atomic qubit) and the qubus. To study the entanglement of such a state and also for later purposes,
it is most convenient to use directly the ${|\widetilde{u}\rangle, |\widetilde{v}\rangle}$-basis on the light mode ,
where $\sim$ refers to the basis vectors in Eqs. \eqref{eq: qubitbasis1} and \eqref{eq: qubitbasis2} with damped amplitudes $\sqrt{\gamma}\alpha$.\\ 
In addition, a basis change
on the matter qubit system into the conjugate \textit{X}-basis, $|\widetilde{0}\rangle=\frac{1}{\sqrt{2}}(|0\rangle+|1\rangle)$
and $|\widetilde{1}\rangle=\frac{1}{\sqrt{2}}(|0\rangle-|1\rangle)$, gives the expression
\begin{equation}
\begin{aligned}
\rho_{out}&=\frac{N_{u}(\sqrt{1-\gamma}\alpha)}{4} \left[\frac{1}{2}\left(\sqrt{N_{u}(\sqrt{\gamma}\alpha)}|\widetilde{0}\rangle|\widetilde{u}\rangle\right.\right.\\
&+\left.\left.\sqrt{N_{v}(\sqrt{\gamma}\alpha)}|\widetilde{1}\rangle|\widetilde{v}\rangle\right)\right]\times H.c.\\
&+\frac{N_{v}(\sqrt{1-\gamma}\alpha)}{4} \left[\frac{1}{2}\left(\sqrt{N_{u}(\sqrt{\gamma}\alpha)}|\widetilde{1}\rangle|\widetilde{u}\rangle\right.\right.\\
&+\left.\left.\sqrt{N_{v}(\sqrt{\gamma}\alpha)}|\widetilde{0}\rangle|\widetilde{v}\rangle)\right)\right]\times H.c.,
\end{aligned}
\end{equation}
 which represents the state in Eq. \eqref{eq: qubitchannel}  in suitable binary orthogonal bases for both the matter system and the qubus.
 Note that this does not change the entanglement properties of the state since 
 any entanglement measure is invariant under local basis changes \cite{Plenio, Guehne, Horodeckireview}. 
 Also note that this matter-light qubit-qubus entangled state effectively remains an entangled qubit-qubit state,
 since the two initial coherent states of the qubus span a two-dimensional qubit space and because 
 individual coherent states remain pure after a loss channel.\\
After traveling through an optical fiber over the distance $L_{0}$, the light mode interacts dispersively with a second 
matter qubit system, also prepared in the state $|0\rangle+|1\rangle$, but this time with the inverse angle, $\theta=-\pi$. \\ 
The joint tripartite state, written in the same basis as in  Eq. \eqref{eq: qubitchannel}, then becomes
\begin{equation}
\begin{aligned}
\label{eq: mixture}
 \rho_{out}&=\frac{N_{u}(\sqrt{1-\gamma}\alpha)}{4} |C_{0}\rangle\langle C_{0}|\\
 &+\frac{N_{v}(\sqrt{1-\gamma}\alpha)}{4} |C_{1}\rangle\langle C_{1}|,
\end{aligned}
 \end{equation}
 where 
 \begin{equation}
  |C_{0}\rangle=\frac{1}{\sqrt{2}}(|\phi^{+}\rangle|\sqrt{\gamma}\alpha\rangle+|\psi^{+}\rangle|-\sqrt{\gamma}\alpha\rangle),
 \end{equation}
and 
\begin{equation}
|C_{1}\rangle=\frac{1}{\sqrt{2}}(|\phi^{-}\rangle|\sqrt{\gamma}\alpha\rangle+|\psi^{-}\rangle|-\sqrt{\gamma}\alpha\rangle).
\end{equation}
Here we introduced the qubit Bell states
\begin{equation}
 \begin{aligned}
  |\phi^{\pm}\rangle&=\frac{1}{\sqrt{2}}(|00\rangle\pm |11\rangle),\\
  |\psi^{\pm}\rangle&=\frac{1}{\sqrt{2}}(|10\rangle\pm|01\rangle).\\
 \end{aligned}
\end{equation}
The component $|C_{0}\rangle$ in Eq. \eqref{eq: mixture} is the desired target component,
whereas $|C_{1}\rangle$ is the loss component that vanishes in the loss-free case. 
Indeed, for $\gamma\rightarrow 1$, one observes $N_{u}(0)=4$ and $N_{v}(0)=0$ such that
in this case the corresponding output density operator $\rho_{out}=|C_{0}\rangle\langle C_{0}|$
represents a pure state. Opposed to the original HQR for qubits \cite{Bright}, here every term
in $|C_{0}\rangle$ contains matter two-qubit entanglement because of our choice $\theta=\pm \pi$. This choice will
enable us later to obtain a natural generalization to qudits.\\
To achieve the goal of distributing entanglement between the two separated matter systems over the distance $L_{0}$, the final step
is a measurement on the light mode, for instance, by homodyne detection.\\
Unlike in the original HQR protocol \cite{Bright} where the dispersive interaction is assumed to be weak
(and hence a \textit{p}-homodyne detection is ultimately preferred over an $x$-homodyne detection with, respectively,
state distinguishabilities $\sim \alpha\theta$ versus $\alpha \theta^{2}$ for small but otherwise unfixed theta),
a suitable detection scheme in our case for "strong" and fixed $\theta=\pm \pi$ is a measurement of the quadrature $\hat{x}=\frac{1}{2}(a+a^{\dagger})$ instead of $\hat{p}=\frac{1}{2i}(a-a^{\dagger})$.\\
The position distribution of coherent states with complex amplitude $\beta$ can be obtained by the square
of their position wave functions,
\begin{equation}
\label{eq: coherent}
|\psi_{\beta}(x)|^{2}=\sqrt{\frac{2}{\pi}}\exp\left(-2(x-\operatorname{Re}(\beta))^{2}\right).
\end{equation}
Because of the finite overlap of the coherent states $|\sqrt{\gamma}\alpha\rangle$ and $|-\sqrt{\gamma}\alpha\rangle$, it is impossible
to perfectly distinguish these states and an error due to this non-orthogonality has to be taken into account.
Based on Eq. \eqref{eq: coherent}, it is obvious that $|\sqrt{\gamma}\alpha\rangle$ and $|-\sqrt{\gamma}\alpha\rangle$
have Gaussian position distributions around $\sqrt{\gamma}\alpha$ and $-\sqrt{\gamma}\alpha$, respectively. It is therefore useful to
assign the result of the \textit{x}-measurement to one of three possible windows.\\
The first window is $w_{0}=[\sqrt{\gamma}\alpha-\Delta,\infty]$ with $\sqrt{\gamma}\alpha>\Delta>0$. 
If the measurement result falls into this range, then the light mode is effectively projected onto 
$|\sqrt{\gamma}\alpha\rangle$. Note that this is an approximate projection due to the non-orthogonality, i.e., the resulting state
is still a superposition of $|\phi^{+}\rangle$ and 
$|\psi^{+}\rangle$ in the first component, while the weight of $|\psi_{+}\rangle$ can be reduced
by increasing the value of $\sqrt{\gamma}\alpha$. The same is true in the second component 
for $|\phi^{-}\rangle$  and $|\psi^{-}\rangle$.\\
As for the second window, we define $w_{1}=[-\infty, -\sqrt{\gamma}\alpha+\Delta]$, which is symmetric to $w_{0}$ and therefore 
represents the approximate projection on $|-\sqrt{\gamma}\alpha\rangle$. Unlike $w_{0}$, one has now
$|\psi^{\pm}\rangle$ as the dominant terms in the superpositions in the two components. It is again true that 
the non-dominant term in the superposition can be made arbitrarily small by increasing $\sqrt{\gamma}\alpha$.
A third window, $w_{2}$, can be defined in between $w_{0}$ and $w_{1}$, and a measurement result in this range will be considered as a failure event
to be discarded (see Fig. \ref{fig: qubitwindows}).\\
Useful figures of merit for the performance of this entanglement distribution scheme are the success probabilities for the 
two non-failure windows $w_{0}$ and $w_{1}$ as well as the fidelity of the corresponding target state in the first component.
As the fidelity, we define the overlap of the maximally entangled Bell states $|\phi^{+}\rangle$ $(w_{0})$
or $|\psi^{+}\rangle$ $(w_{1})$ with the mixed state in Eq. \eqref{eq: mixture} after the corresponding homodyne measurement outcome.\\
The success probability for a measurement result to fall into the first window reads
\begin{equation}
p_{w_{0}}=\frac{1}{2}\int\limits_{\sqrt{\gamma}\alpha-\Delta}^{\infty}dx(|\psi_{\sqrt{\gamma}\alpha}(x)|^{2}+|\psi_{-\sqrt{\gamma}\alpha}(x)|^{2}).
\end{equation}
For the second window, we have 
\begin{equation}
 p_{w_{1}}=\frac{1}{2}\int\limits_{-\infty}^{-\sqrt{\gamma}\alpha+\Delta}dx(|\psi_{\sqrt{\gamma}\alpha}(x)|^{2}+|\psi_{-\sqrt{\gamma}\alpha}(x)|^{2}),
 \end{equation}
 which equals $p_{w_{0}}$ for symmetry reasons. The same holds true for the two fidelities,
 \begin{equation}
 \begin{aligned}
  F_{w_{0}}&=F_{w_{1}}\\
  &=\frac{N_{u}(\sqrt{1-\gamma}\alpha)}{4}\\
  &\times\frac{\int\limits_{-\infty}^{-\sqrt{\gamma}\alpha+\Delta}dx|\psi_{\sqrt{\gamma}\alpha}(x)|^{2}}{\int\limits_{-\infty}^{-\sqrt{\gamma}\alpha+\Delta}dx(|\psi_{\sqrt{\gamma}\alpha}(x)|^{2}+|\psi_{-\sqrt{\gamma}\alpha}(x)|^{2})}.
\end{aligned}
  \end{equation}
The formulae for the fidelities and the success probabilities imply the crucial
dependence of the performance on the choice of $\Delta$ and $\sqrt{\gamma}\alpha$:
if we choose $\Delta=\Delta_{0}:=\sqrt{\gamma}\alpha$, then we have no failure window and every
measurement result is assigned to one of the two coherent states $|\pm\sqrt{\gamma}\alpha\rangle$.
The corresponding success probability equals unity at the expense of a rather low fidelity.\\
With $\Delta<\Delta_{0}$, the success probability is clearly less than unity and the fidelity
increases correspondingly.\\
In general, the fidelity drops for too small $\sqrt{\gamma}\alpha$ due to 
the non-orthogonality and thus indistinguishability of the coherent states 
$|\pm\sqrt{\gamma}\alpha\rangle$. The overall effect becomes manifest in bit-flip errors
in the target Bell states.
Though leading to near-orthogonality, large amplitudes $\sqrt{\gamma}\alpha$
result in a near-equal mixture of the state in Eq. \eqref{eq: mixture}
which then, after a near-deterministic discrimination, consists of one of the two possible Bell states in the first component and its 
phase-flipped version in the second component. This state  therefore has very low entanglement and hence is of limited practical interest.
So the task is to find a regime of $\alpha$ and
distances $L_{0}$ such that both reasonable fidelities and success probabilities can be obtained.\\
Besides homodyne detection, unambiguous state discrimination (USD) has been considered for hybrid quantum repeaters in the 
literature \cite{HQR}. The advantage here is that the effects originating from the finite overlaps 
of the coherent states no longer appear in the fidelity thanks to an error-free
state discrimination. The corresponding effects solely influence the success probabilities depending on 
the weights of the inconclusive discrimination results.
Two-state USD for coherent states $|\pm \sqrt{\gamma}\alpha\rangle$ is well-known
and can be optimally performed via a single beam splitter and on-off detections \cite{BanaszekUSD}.\\
Further steps in the original repeater protocol address the purification of the mixed state in Eq. \eqref{eq: mixture} after homodyne detection
and entanglement swapping on the matter system or via the qubus to distribute the generated entanglement over longer distances.\\
For more details, see e.g. \cite{HQR}.
\section{Hybrid quantum repeater for qutrits}
\label{sec: qutrit}
\subsection{Dispersive light-matter interaction}
The dispersive interaction (see Eq. \eqref{eq: qubitdis}) lies at the heart of the HQR
for qubits and therefore, as a first step to extend this repeater scheme to qutrits, a generalization 
of the dispersive interaction to the qutrit case is necessary.\\
In analogy to the dispersive interaction for qubits, we define the qutrit-qubus interaction Hamiltonian 
as
\begin{equation}
 H_{int}^{(3)}=\hbar g S_{z}^{(3)}a^{\dagger}a,
\end{equation}
where the operator $S_{z}^{(3)}$ acts on the qutrit basis states  $|0\rangle$, $|1\rangle$ and $|2\rangle$ as
\begin{equation}
\label{eq: qutritz}
 \begin{aligned}
  S_{z}^{(3)}|0\rangle&=-1\cdot |0\rangle,\\
  S_{z}^{(3)}|1\rangle&=0\cdot |1\rangle,\\
  S_{z}^{(3)}|2\rangle&=1\cdot |2\rangle.\\
 \end{aligned}
\end{equation}
The matter system could be, for example, realized by a spin-1 particle where the basis states
are the eigenstates with the corresponding magnetic quantum numbers, $m_{z}=-1,0,1$. Such a spin realization of a qutrit
has been demonstrated in the framework of nuclear magnetic resonance (NMR) for various applications \cite{NMR1,NMR2}.\\ 
Similar to the qubit case, the corresponding unitary transformation is
$U_{3}(\theta)=\exp\left(i\theta S_{z}^{(3)}a^{\dagger}a\right)$, which again 
corresponds to a conditional phase rotation on the light-matter system
(up to an unconditional phase shift of the qubus mode by $e^{i\theta}$), i.e.,
\begin{equation}
 U_{3}(\theta)(|0\rangle+|1\rangle+|2\rangle)\otimes|\alpha\rangle
 =|0\rangle|\alpha\rangle+|1\rangle|\alpha e^{i\theta}\rangle+|2\rangle|\alpha e^{2i\theta}\rangle.
\end{equation}
For our purposes, we will choose $\theta=\frac{2\pi}{3}$ to obtain a rather strong dispersive interaction.
\subsection{Loss-free case}
The qutrit hybrid repeater protocol works in complete analogy to the qubit case.
To illustrate the concept, we first omit photon losses in the optical fiber and assume
a noiseless quantum channel.\\
The repeater protocol works as follows:
First, the matter system is initiated in the state $\frac{1}{\sqrt{3}}(|0\rangle+|1\rangle+|2\rangle)$
and interacts with a light mode in a coherent state $|\alpha\rangle$ via the qutrit dispersive interaction
with $\theta=\frac{2\pi}{3}$. This results in the entangled matter-qubus state
\begin{equation}
\label{eq: qutritprep}
 \frac{1}{\sqrt{3}}\left(|0\rangle|\alpha\rangle+|1\rangle|\alpha e^{\frac{2\pi i}{3}}\rangle+|2\rangle|\alpha e^{-\frac{2\pi i}{3}}\rangle\right).
\end{equation}
The light mode is then sent to a second matter system, separated from the first one 
by a distance $L_{0}$ and also prepared in the state $\frac{1}{\sqrt{3}}(|0\rangle+|1\rangle+|2\rangle)$.
The incoming light mode interacts dispersively with the second matter system, but this time with the reverse angle
$\theta=-\frac{2\pi}{3}$. The resulting pure state is
\begin{equation}
 \begin{aligned}
 \label{eq: noloss}
\frac{1}{\sqrt{3}}&\left(\frac{1}{\sqrt{3}}(|00\rangle+|11\rangle+|22\rangle)|\alpha\rangle\right.\\
 &+\frac{1}{\sqrt{3}}(|02\rangle+|10\rangle+|21\rangle)|\alpha e^{\frac{2\pi i}{3}}\rangle\\
 &+\left. \frac{1}{\sqrt{3}}(|01\rangle+|12\rangle+|20\rangle)|\alpha e^{-\frac{2\pi i}{3}}\rangle\right).
 \end{aligned}
 \end{equation}
To keep the notation short and also for later purposes, it it useful
to define the set of maximally entangled qutrit Bell states, 
\begin{equation}
\label{eq: qutritBell}
 |\phi_{kj}\rangle=\frac{1}{\sqrt{3}}\sum\limits_{m=0}^{2}\exp\left(\frac{2\pi i k m}{3}\right)|m,m\ominus j\rangle,
\end{equation}
where $"\ominus"$ denotes subtraction modulo 2. Eq. \eqref{eq: noloss} can therefore be rewritten
as
\begin{equation}
 \frac{1}{\sqrt{3}}\left(|\phi_{00}\rangle|\alpha\rangle
 + |\phi_{01}\rangle|\alpha e^{\frac{2\pi i}{3}}\rangle
 + |\phi_{02}\rangle|\alpha e^{-\frac{2\pi i}{3}}\rangle
 \right).
\end{equation}

To generate a maximally entangled state between the matter systems, a homodyne measurement is performed 
on the light mode to distinguish the three coherent states of the mode. 
Unlike the qubit case, here a measurement of $\hat{p}$ is useful, because it allows one to (almost)
discriminate all three coherent states (as opposed to the case of an $\hat{x}$-measurement).
Moreover, for an ideal loss-free channel, increasing the amplitude $\alpha$ leads to near-orthogonality of the coherent states
such that a perfect, near-maximally entangled qutrit-qutrit state can be deterministically distributed over the distance $L_{0}$.
To further extend the entanglement, two such elementary pairs next to each other are connected by entanglement swapping,
via a Bell measurement on adjacent repeater nodes. By one successful entanglement
swapping step, qutrit-qutrit entanglement can thus be shared over the distance $2L_{0}$, and so forth.
\\
We will address all the steps of the qutrit repeater protocol in detail in the next sections and also explain which subtleties
and necessary generalizations occur in practice compared to the idealized loss-free case discussed here.
\subsection{Matter-light qutrit-qubus hybrid entanglement}
\begin{figure}[t!]
\centering
\includegraphics[width=0.55\textwidth]{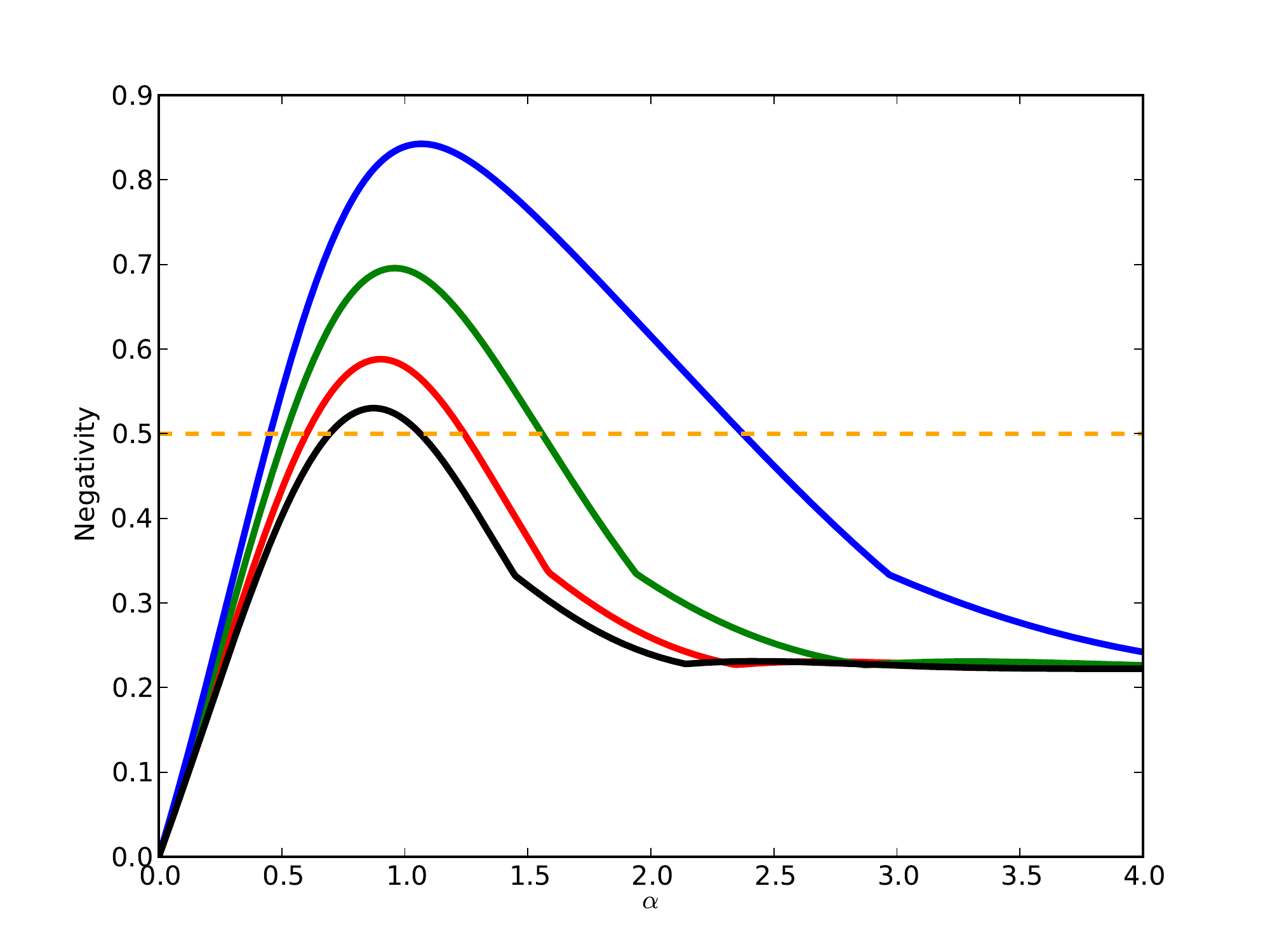}
\caption{Negativity of the effective qutrit-qutrit state in dependence of $\alpha$ for various distances: 10 km (black), 8 km (red), 5 km (green) and 2 km (blue) (from bottom to top).
The dashed, orange line indicates the negativity of a maximally entangled pure two-qubit Bell state.}
\label{fig: qutritnegativity}
\end{figure}

At the beginning of the qutrit HQR protocol, the matter system is prepared
in the state $\frac{1}{\sqrt{3}}(|0\rangle+|1\rangle+|2\rangle)$. The dispersive interaction with a coherent state leads to
the state in Eq. \eqref{eq: qutritprep}.
In the realistic case,  the light mode is sent through an optical loss channel (e.g. an optical fiber), which is again simulated by a coupling
of the mode with an ancilla vacuum state. This time,  the application of the beam splitter leads to
\begin{equation}
\label{eq: qutritbeam}
\begin{aligned}
 \frac{1}{\sqrt{3}}(&|0\rangle|\sqrt{\gamma}\alpha\rangle|\sqrt{1-\gamma}\alpha\rangle+
 |1\rangle|\sqrt{\gamma}\alpha e^{\frac{2\pi i}{3}}\rangle|\sqrt{1-\gamma}\alpha e^{\frac{2\pi i}{3}}\rangle\\
 +&|2\rangle|\sqrt{\gamma}\alpha e^{-\frac{2\pi i}{3}}\rangle|\sqrt{1-\gamma}\alpha e^{-\frac{2\pi i}{3}}\rangle).
 \end{aligned}
 \end{equation}
 To trace out the loss mode, it is again useful to switch to an orthogonal basis. While in the qubit case that basis
 is given by a kind of qubit Hadamard transform, the qutrit basis is given by a qutrit Hadamard gate to yield
\begin{equation}
\label{eq: qutritbasis}
 \begin{aligned}
  |u\rangle&=\frac{1}{\sqrt{N_{u}(\alpha)}}(|\alpha\rangle+|\alpha e^{\frac{2\pi i}{3}}\rangle+|\alpha e^{-\frac{2\pi i}{3}}\rangle),\\
  |v\rangle&=\frac{1}{\sqrt{N_{v}(\alpha)}}(|\alpha\rangle+e^{\frac{2\pi i}{3}}|\alpha e^{\frac{2\pi i}{3}}\rangle+e^{-\frac{2\pi i}{3}}|\alpha e^{-\frac{2\pi i}{3}}\rangle),\\
 |w\rangle&=\frac{1}{\sqrt{N_{w}(\alpha)}}(|\alpha\rangle+e^{-\frac{2\pi i}{3}}|\alpha e^{\frac{2\pi i}{3}}\rangle+e^{\frac{2\pi i}{3}}|\alpha e^{-\frac{2\pi i}{3}}\rangle),\\
 \end{aligned}
 \end{equation}
 with normalization constants
 \begin{equation}
 \begin{aligned}
 N_{u}(\alpha)&=3+6e^{-\frac{3}{2}\alpha^{2}}\cos\left(\sqrt{\frac{3}{4}}\alpha^{2}\right),\\
 N_{v}(\alpha)&=3-e^{-\frac{3}{2}\alpha^{2}}\left(3\cos\left(\sqrt{\frac{3}{4}}\alpha^{2}\right)\right.\\
 &+\left.\sqrt{3}\sin\left(\sqrt{\frac{3}{4}}\alpha^{2}\right)\right),\\
 N_{w}(\alpha)&=3-e^{-\frac{3}{2}\alpha^{2}}\left(3\cos\left(\sqrt{\frac{3}{4}}\alpha^{2}\right)\right.\\
 -&\left.\sqrt{3}\sin\left(\sqrt{\frac{3}{4}}\alpha^{2}\right)\right).\\
 \end{aligned}
 \end{equation}
 The coherent states above can thus be written as
 \begin{equation}
 \begin{aligned}
 |\alpha\rangle&=\frac{1}{3}(\sqrt{N_{u}(\alpha)}|u\rangle+\sqrt{N_{v}(\alpha)}|v\rangle+\sqrt{N_{w}(\alpha)}|w\rangle),\\
 |\alpha e^{\frac{2\pi i}{3}}\rangle&=\frac{1}{3}(\sqrt{N_{u}(\alpha)}|u\rangle+e^{-\frac{2\pi i}{3}}\sqrt{N_{v}(\alpha)}|v\rangle\\
 &+e^{\frac{2\pi i}{3}}\sqrt{N_{w}(\alpha)}|w\rangle),\\
  |\alpha e^{-\frac{2\pi i}{3}}\rangle&=\frac{1}{3}(\sqrt{N_{u}(\alpha)}|u\rangle+e^{\frac{2\pi i}{3}}\sqrt{N_{v}(\alpha)}|v\rangle\\
  &+e^{-\frac{2\pi i}{3}}\sqrt{N_{w}(\alpha)}|w\rangle).\\
 \end{aligned}
 \end{equation}
 Substituting this into Eq. \eqref{eq: qutritbeam} for the loss mode and tracing out the loss mode gives the three-component mixed state
 \begin{equation}
 \label{eq: qutritchannel}
 \begin{aligned}
 \rho_{out}&=\frac{N_{u}(\sqrt{1-\gamma}\alpha)}{9} \left[\frac{1}{\sqrt{3}}(|0\rangle|\sqrt{\gamma}\alpha\rangle+|1\rangle|\sqrt{\gamma}\alpha e^{\frac{2\pi i}{3}}\rangle\right.\\
 &+\left.|2\rangle|\sqrt{\gamma}\alpha e^{-\frac{2\pi i}{3}}\rangle)\right]\times H.c.\\
&+\frac{N_{v}(\sqrt{1-\gamma}\alpha)}{9} \left[\frac{1}{\sqrt{3}}(|0\rangle|\sqrt{\gamma}\alpha\rangle+e^{-\frac{2\pi i}{3}}|1\rangle|\sqrt{\gamma}\alpha e^{\frac{2\pi i}{3}}\rangle\right.\\
& +\left.e^{\frac{2\pi i}{3}}|2\rangle|\sqrt{\gamma}\alpha e^{\frac{-2\pi i}{3}}\rangle)\right]\times H.c.\\
 &+\frac{N_{w}(\sqrt{1-\gamma}\alpha)}{9} \left[\frac{1}{\sqrt{3}}(|0\rangle|\sqrt{\gamma}\alpha\rangle+e^{\frac{2\pi i}{3}}|1\rangle|\sqrt{\gamma}\alpha e^{\frac{2\pi i}{3}}\rangle\right.\\
 &+\left.e^{-\frac{2\pi i}{3}}|2\rangle|\sqrt{\gamma}\alpha e^{-\frac{2\pi i}{3}}\rangle)\right]\times H.c.\\
 \end{aligned}
 \end{equation}
 This represents an entangled state between the qutrit matter system and the qubus.
 Similar to the qubit case, the resulting density matrix still effectively represents
 a state of two qutrits (one optical and one material), since
 the three coherent states $\{|\sqrt{\gamma}\alpha\rangle, |\sqrt{\gamma}\alpha e^{\pm \frac{2\pi i}{3}}\rangle\}$
 effectively span a three-dimensional Hilbert space.\\
 For studying the entanglement properties of $\rho_{out}$,
 it is helpful to express the light mode in the $\{|u\rangle,|v\rangle,|w\rangle\}$- basis and the matter system 
 in the qutrit (generalized Pauli) \textit{X}-basis,
 \begin{equation}
 \begin{aligned}
 |\widetilde{0}\rangle&=\frac{1}{\sqrt{3}}(|0\rangle+|1\rangle+|2\rangle),\\
 |\widetilde{1}\rangle&=\frac{1}{\sqrt{3}}(|0\rangle+e^{\frac{2\pi i}{3}}|1\rangle+e^{-\frac{2\pi i}{3}}|2\rangle),\\
 |\widetilde{2}\rangle&=\frac{1}{\sqrt{3}}(|0\rangle+e^{-\frac{2\pi i}{3}}|1\rangle+e^{\frac{2\pi i}{3}}|2\rangle).
 \end{aligned}
 \end{equation}
 Eq. \eqref{eq: qutritchannel} can thus be rewritten as
 \begin{equation}
 \begin{aligned}
  &\rho_{out}=\frac{N_{u}(\sqrt{1-\gamma}\alpha)}{9} \left[\frac{1}{3}\left(\sqrt{N_{u}(\sqrt{\gamma}\alpha)}|\widetilde{0}\rangle |\widetilde{u}\rangle\right.\right.\\
  &+\left.\left.\sqrt{N_{v}(\sqrt{\gamma}\alpha)}|\widetilde{1}\rangle|\widetilde{v}\rangle
 +\sqrt{N_{w}(\sqrt{\gamma}\alpha)}|\widetilde{2}\rangle|\widetilde{w}\rangle\right)\right]\times H.c.\\
 &+\frac{N_{v}(\sqrt{1-\gamma}\alpha)}{9} \left[\frac{1}{3}\left(\sqrt{N_{u}(\sqrt{\gamma}\alpha)}|\widetilde{2}\rangle |\widetilde{u}\rangle\right.\right.\\
 &+\left.\sqrt{N_{v}(\sqrt{\gamma}\alpha)}|\widetilde{1}\rangle|\widetilde{v}\rangle
 +\left.\sqrt{N_{w}(\sqrt{\gamma}\alpha)}|\widetilde{0}\rangle|\widetilde{w}\rangle\right)\right]\times H.c.\\
 &+\frac{N_{w}(\sqrt{1-\gamma}\alpha)}{9} \left[\frac{1}{3}\left(\sqrt{N_{u}(\sqrt{\gamma}\alpha)}|\widetilde{1}\rangle |\widetilde{u}\rangle\right.\right.\\
 &+\sqrt{N_{v}(\sqrt{\gamma}\alpha)}|\widetilde{0}\rangle|\widetilde{v}\rangle
 +\left.\left.\sqrt{N_{w}(\sqrt{\gamma}\alpha)}|\widetilde{2}\rangle|\widetilde{w}\rangle\right)\right]\times H.c.~ ,\\
 \end{aligned}
 \end{equation}
 where $|\widetilde{u}\rangle, |\widetilde{v}\rangle$ and $|\widetilde{w}\rangle$
 denote the basis vectors in Eq. \eqref{eq: qutritbasis} with amplitudes $\sqrt{\gamma}\alpha$.\\
 To quantify the qutrit-qutrit entanglement of this state, we choose the so-called entanglement negativity \cite{Werner, PlenioPRL}
 as our figure of merit. The negativity $\mathcal{N}$ of a bipartite quantum state of a system $AB$ is defined as
 \begin{equation}
  \mathcal{N}(\rho)=\frac{||\rho^{T_{A}}||-1}{2},
 \end{equation}
where $\rho^{T_{A}}$ is the partial transposition of the bipartite state with respect to system $A$ 
and $||\circ||$ denotes the trace norm.\\
A plot of the negativities for different initial amplitudes $\alpha$ and various elementary distances 
$L_{0}$ with $\gamma=\exp\left(-\frac{L_{0}}{L_{att}}\right)$ is shown in Fig.\ref{fig: qutritnegativity}.
The dashed orange line  indicates the entanglement negativity of a pure maximally entangled qubit Bell state. Up to 
a distance of approximately $L_{0}=10~\text{km}$, it is possible to generate matter-qubus entanglement stronger than 
any, even ideal qubit-qubit entanglement. Taking into account that the realistic distribution of qubit-qubit entanglement is also subject
to loss, the difference in entanglement negativity will be even more significant. However, a crucial step still is to transfer
this entanglement to a sufficient extend from the matter-light system to a matter-matter system for storage.
\subsection{Matter-matter qutrit-qutrit entanglement}
\label{sec: matter-matter}
\begin{figure}[t!]
\centering
\includegraphics[width=0.45\textwidth]{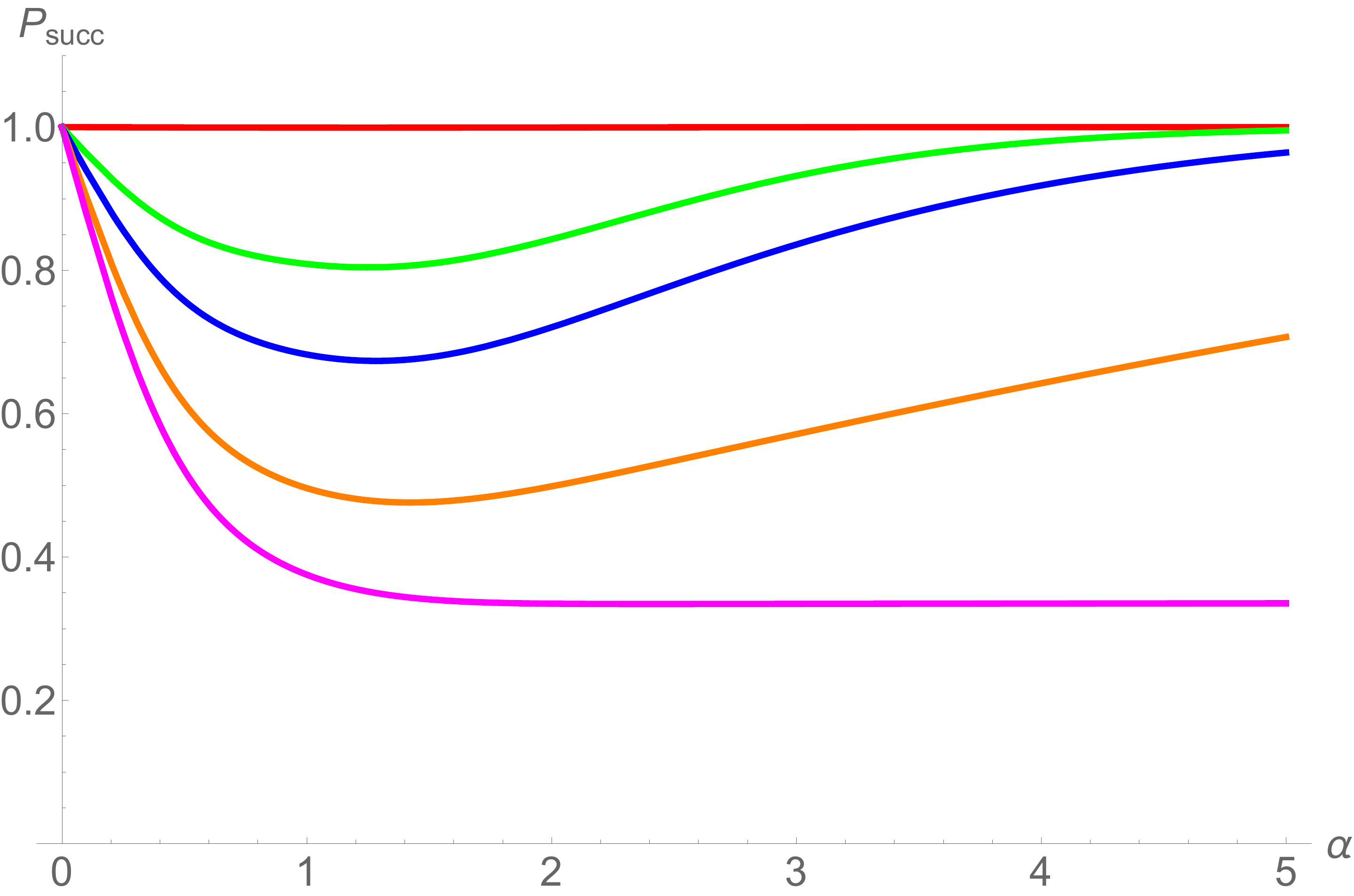}
\caption{Success probability for the homodyne-based distribution of
qutrit-qutrit entanglement over a distance of 5 km for various $\Delta$:
$\Delta=\Delta_{0}$ (red), $\Delta=0.7\Delta_{0}$ (green),
$\Delta=0.5\Delta_{0}$ (blue), $\Delta=0.2\Delta_{0}$ (orange) and
$\Delta=0.001\Delta_{0}$ (magenta) (from top to bottom).
}
\label{fig: qutritprob}
\end{figure}

\begin{figure}[t!]
\centering
\includegraphics[width=0.45\textwidth]{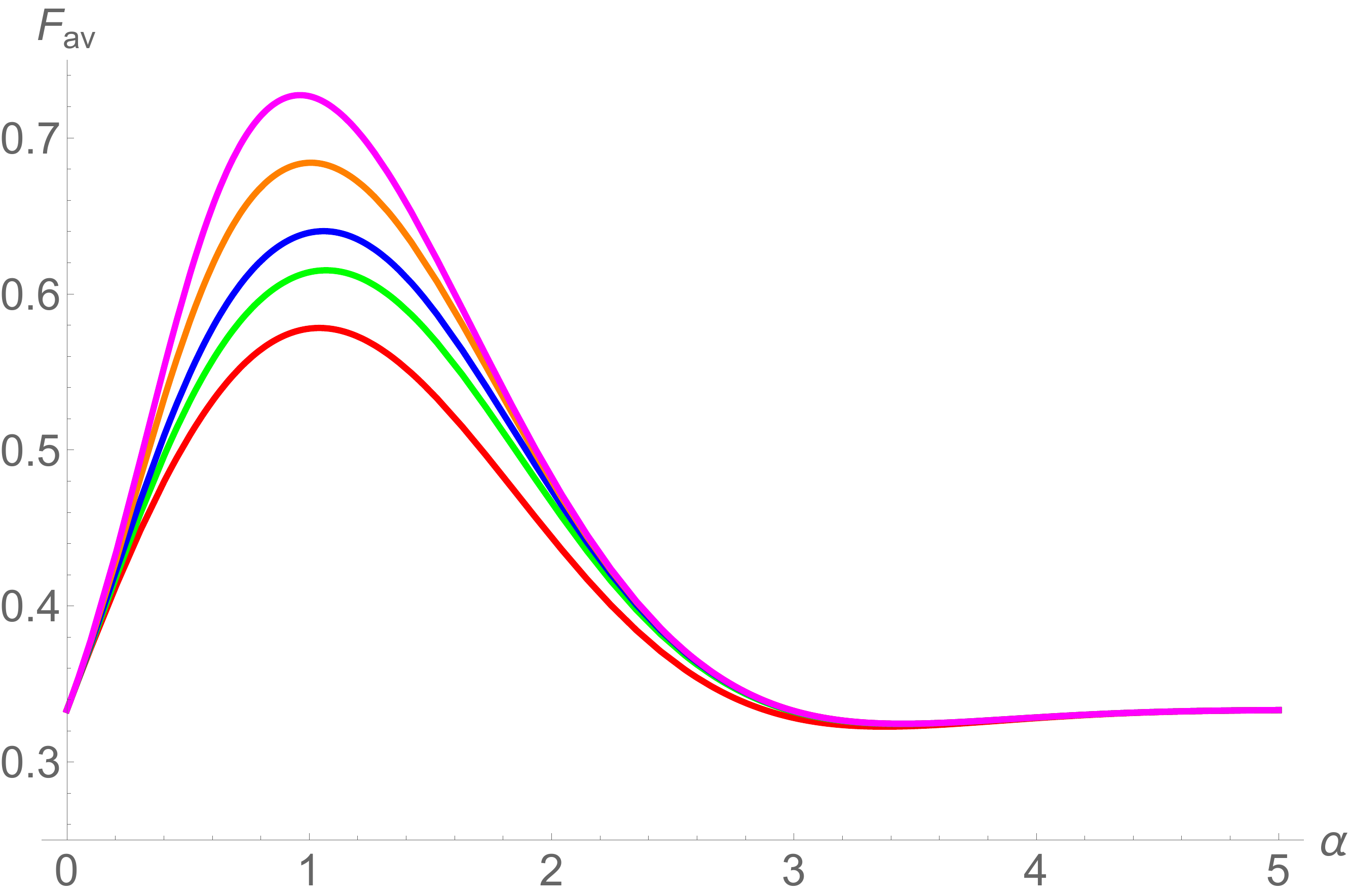}
\caption{Average fidelity for the homodyne-based distribution of
qutrit-qutrit entanglement over a distance of 5 km for various $\Delta$:
$\Delta=\Delta_{0}$ (red), $\Delta=0.7\Delta_{0}$ (green),
$\Delta=0.5\Delta_{0}$ (blue), $\Delta=0.2\Delta_{0}$ (orange) and
$\Delta=0.001\Delta_{0}$ (magenta) (from bottom to top).
}
\label{fig: qutritfid}
\end{figure}

\begin{figure}[t!]
\centering
\includegraphics[width=0.45\textwidth]{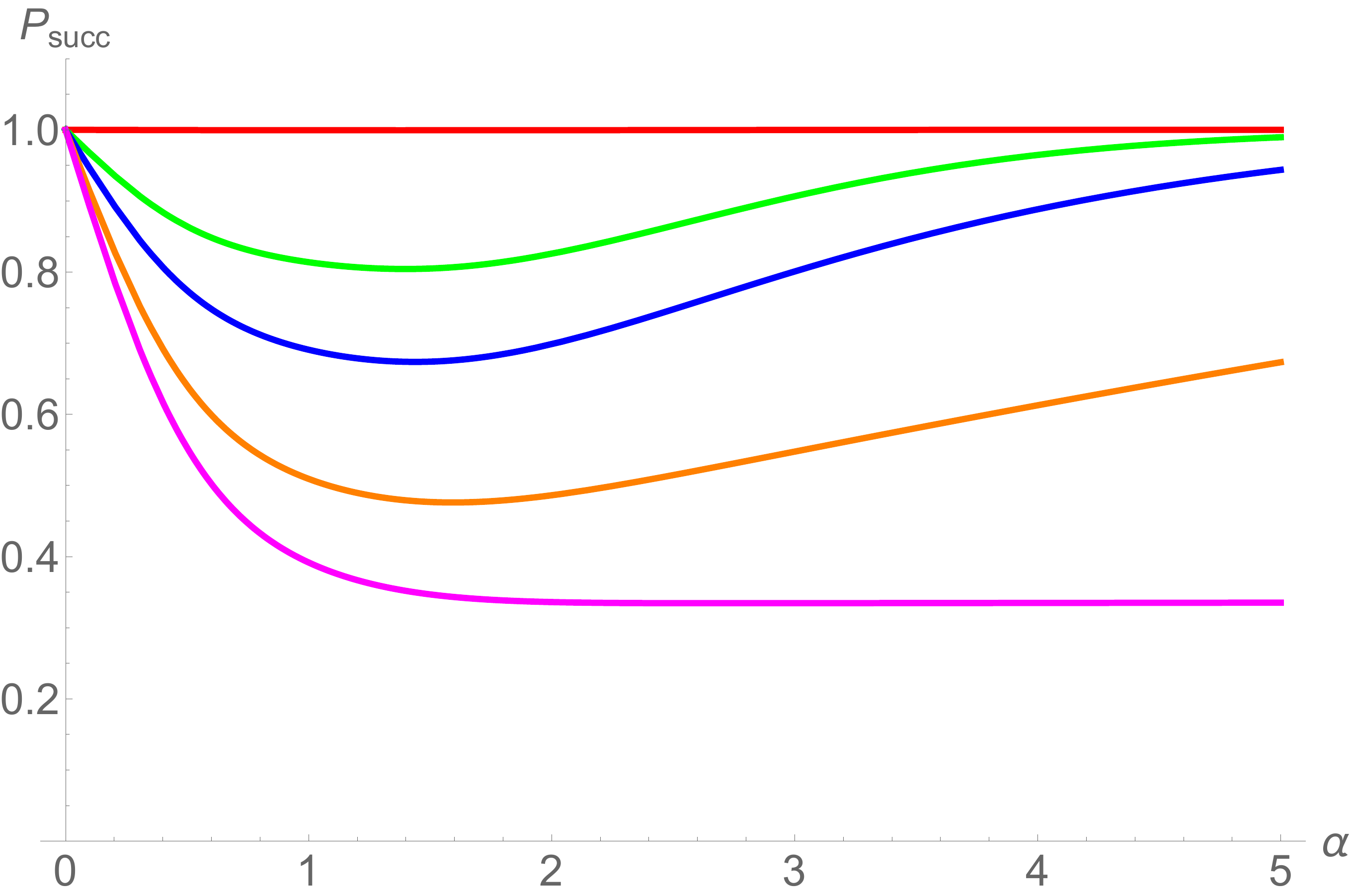}
\caption{Success probability for the homodyne-based distribution of
qutrit-qutrit entanglement over a distance of 10 km for various $\Delta$:
$\Delta=\Delta_{0}$ (red), $\Delta=0.7\Delta_{0}$ (green),
$\Delta=0.5\Delta_{0}$ (blue), $\Delta=0.2\Delta_{0}$ (orange) and
$\Delta=0.001\Delta_{0}$ (magenta) (from top to bottom).
}
\label{fig: qutritprob2}
\end{figure}

\begin{figure}[t!]
\centering
\includegraphics[width=0.45\textwidth]{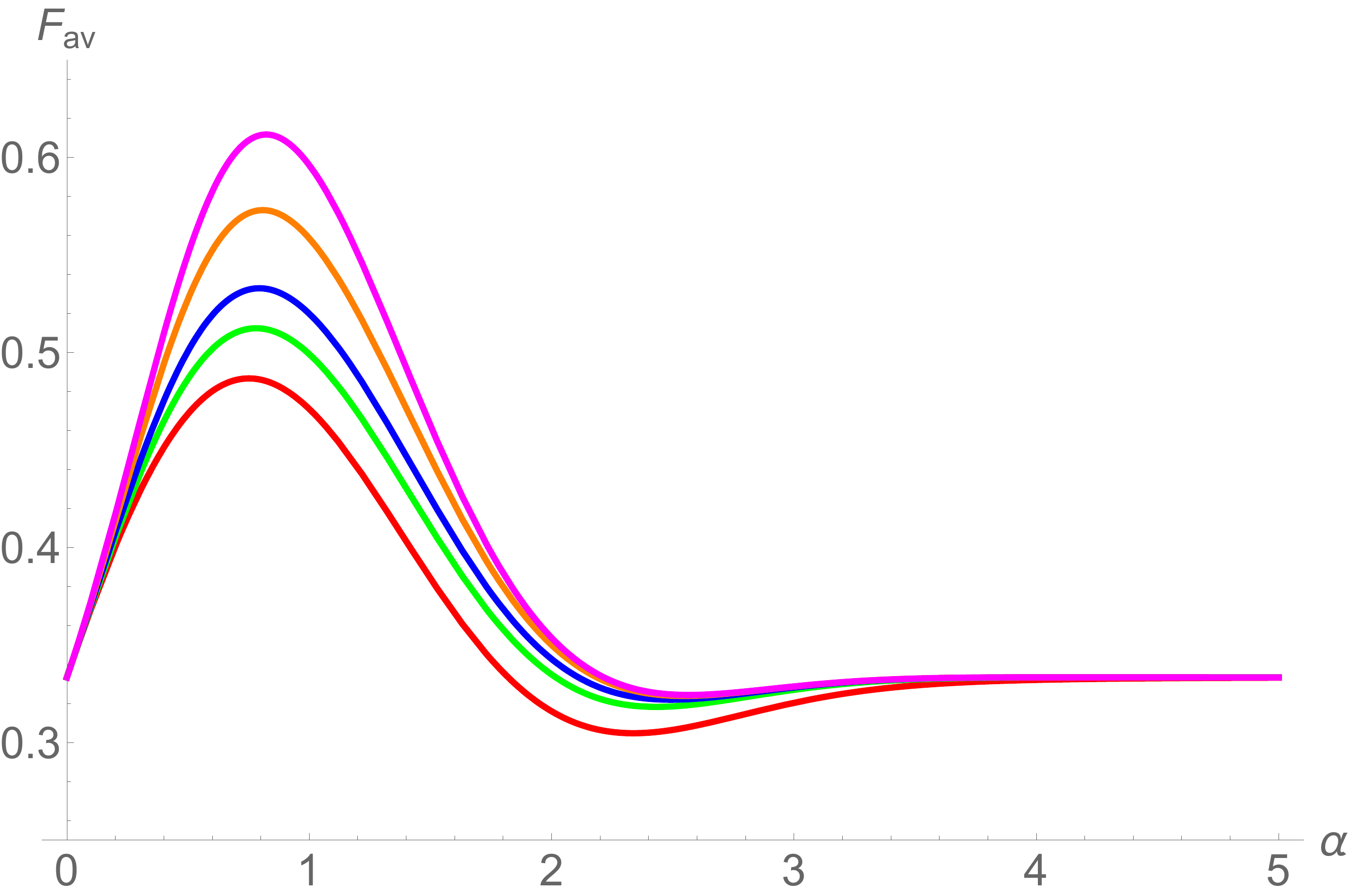}
\caption{Average fidelity for the homodyne-based distribution of
qutrit-qutrit entanglement over a distance of 10 km for various $\Delta$:
$\Delta=\Delta_{0}$ (red), $\Delta=0.7\Delta_{0}$ (green),
$\Delta=0.5\Delta_{0}$ (blue), $\Delta=0.2\Delta_{0}$ (orange) and
$\Delta=0.001\Delta_{0}$ (magenta) (from bottom to top).
}
\label{fig: qutritfid2}
\end{figure}

To distribute entanglement between two matter qutrits, the light mode 
of the state in Eq.\eqref{eq: qutritchannel} interacts with
a second matter system, initialized in the state $\frac{1}{\sqrt{3}}(|0\rangle+|1\rangle+|2\rangle)$.
This time, similar to the qubit case, the controlled phase rotation takes place with the opposite angle, $\theta=-\frac{2\pi}{3}$.
One obtains
\begin{equation}
\label{eq: qutritout}
 \begin{aligned}
  \rho_{out}&=\frac{N_{u}(\sqrt{1-\gamma}\alpha)}{9}|C_{0}\rangle\langle C_{0}|\\
  &+\frac{N_{v}(\sqrt{1-\gamma}\alpha)}{9}|C_{1}\rangle\langle C_{1}|\\
  &+\frac{N_{w}(\sqrt{1-\gamma}\alpha)}{9}|C_{2}\rangle\langle C_{2}|,
 \end{aligned}
\end{equation}
where the individual components are given by
\begin{equation}
 \begin{aligned}
  |C_{0}\rangle&=\frac{1}{\sqrt{3}}(|\phi_{00}\rangle|\sqrt{\gamma}\alpha\rangle\\
 &+ |\phi_{02}\rangle|\sqrt{\gamma}\alpha e^{-\frac{2\pi i}{3}}\rangle
 + |\phi_{01}\rangle|\sqrt{\gamma}\alpha e^{\frac{2\pi i}{3}}\rangle),
 \end{aligned}
 \end{equation}
 \begin{equation}
 \begin{aligned}
 |C_{1}\rangle&=\frac{1}{\sqrt{3}}(|\phi_{20}\rangle|\sqrt{\gamma}\alpha\rangle
 +|\phi_{22}\rangle|\sqrt{\gamma}\alpha e^{-\frac{2\pi i}{3}}\rangle\\ 
 &+|\phi_{21}\rangle|\sqrt{\gamma}\alpha e^{\frac{2\pi i}{3}}\rangle),
 \end{aligned}
\end{equation}

\begin{equation}
\begin{aligned}
|C_{2}\rangle&=\frac{1}{\sqrt{3}}(|\phi_{10}\rangle|\sqrt{\gamma}\alpha\rangle+|\phi_{12}\rangle|\sqrt{\gamma}\alpha e^{-\frac{2\pi i}{3}}\rangle\\
&+|\phi_{11}\rangle|\sqrt{\gamma}\alpha e^{\frac{2\pi i}{3}}\rangle),
 \end{aligned}
 \end{equation}
 with the two-qutrit Bell states from Eq.\eqref{eq: qutritBell}.
 \begin{figure}[t!]
\centering
\includegraphics[width=0.5\textwidth]{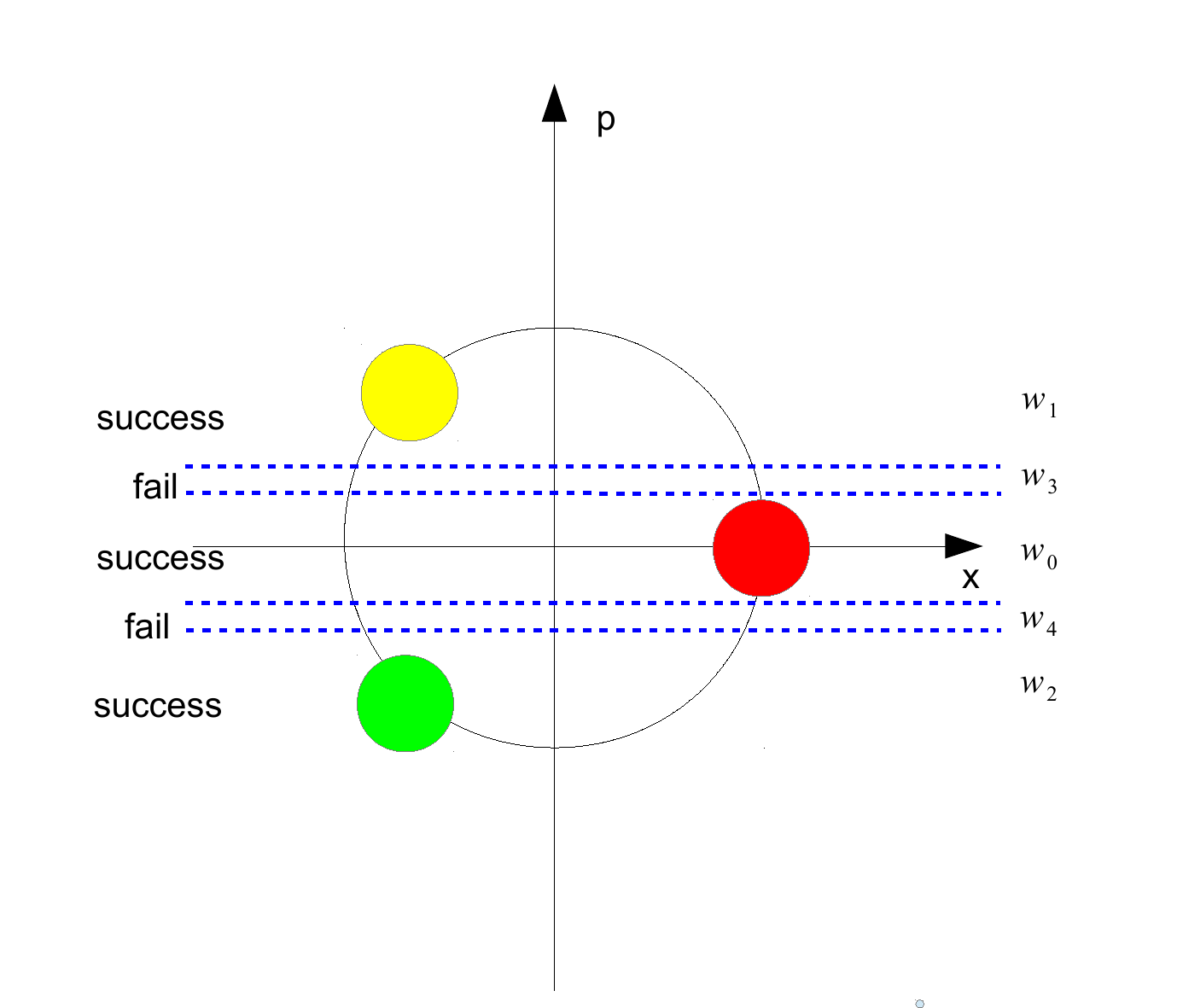}
\caption{Phase space representation of the three coherent states
$|\alpha\rangle$ and $|\alpha e^{\pm\frac{2\pi i}{3}}\rangle$ to be distinguished
by USD.}
\label{fig: qutritpic}
\end{figure}
\noindent
In order to obtain entanglement between the two matter systems, 
 the coherent states $|\sqrt{\gamma}\alpha\rangle,|\sqrt{\gamma}\alpha e^{-\frac{2\pi i}{3}}\rangle$, and $|\sqrt{\gamma}\alpha e^{\frac{2\pi i}{3}}\rangle$
 have to be distinguished (see Fig. \ref{fig: qutritpic}). Like in the loss-free case, this can be done using a homodyne measurement on the light mode.
 Unlike the qubit case, an $\hat{x}$-measurement is not suitable here, because $|\sqrt{\gamma}\alpha e^{\frac{2\pi i}{3}}\rangle$
 and $|\sqrt{\gamma}\alpha e^{-\frac{2\pi i}{3}}\rangle$ cannot be distinguished. Therefore, we choose
 the quadrature $\hat{p}$ whose Gaussian momentum distribution
 for coherent states with complex amplitude $\beta$
 reads as
 \begin{equation}
|\psi_{\beta}(p)|^{2}=\sqrt{\frac{2}{\pi}}\exp\left(-2(p-\operatorname{Im}(\beta))^{2}\right).
\end{equation}
\noindent
This time, it is useful to define at least three windows to which a measurement result is assigned when
the light mode of the output state in Eq.\eqref{eq: qutritout} is measured (see Fig.\ref{fig: qutritpic}).
The first window is a symmetric interval around $p=0$, $w_{0}=[-\Delta,\Delta]$. A measurement
result in this interval, similar to the qubit case, corresponds to an approximate 
projection on $|\alpha\rangle$.
A projection onto the states $|\sqrt{\gamma}\alpha e^{\pm\frac{2\pi i}{3}}\rangle$
is assumed if a value falls into $w_{1}=[ \frac{\sqrt{3}}{2}\sqrt{\gamma}\alpha-\Delta, \infty]$ or
$w_{2}=[ -\infty,-\frac{\sqrt{3}}{2}\sqrt{\gamma}\alpha+\Delta]$, respectively. Note that we need 
$\Delta\leq\frac{1}{2}\sqrt{\frac{3}{4}}\sqrt{\gamma}\alpha=:\Delta_{0}$ to exclude overlapping windows.
We may decide to add two extra windows $w_{3}$
and $w_{4}$ to include the possibility of discarding measurement results (see Fig. \ref{fig: qutritpic}).
Inclusion of such failure events renders our qutrit entanglement distribution probabilistic.\\
Using the momentum wave functions for the coherent states, 
the qutrit-qutrit-qubus $|C_{0}\rangle$-component
of $\rho_{out}$ after measuring the value $p$
in the homodyne detection of the qubus
has the following conditional state for the two
matter qutrits,
\begin{equation}
\label{eq: qutritp}
 \begin{aligned}
 \sigma_{p}^{C_{0}}&= \text{Tr}_{qubus}(|p\rangle\langle p|C_{0}\rangle\langle C_{0}|p\rangle\langle p|)\\
  &=\frac{1}{3}\bigg(|\phi_{00}\rangle\langle\phi_{00}|\cdot |\psi_{\sqrt{\gamma}\alpha}(p)|^{2}\\
  &+|\phi_{02}\rangle\langle\phi_{02}|\cdot |\psi_{\sqrt{\gamma}\alpha e^{-\frac{2\pi i}{3}} }(p)|^{2}\\
  &+|\phi_{01}\rangle\langle\phi_{01}|\cdot |\psi_{\sqrt{\gamma}\alpha e^{\frac{2\pi i}{3}} }(p)|^{2}\\
  &+|\phi_{00}\rangle\langle\phi_{02}|\cdot\psi_{\sqrt{\gamma}\alpha}(p)\psi^{*}_{\sqrt{\gamma}\alpha e^{-\frac{2\pi i}{3}} }(p)\\
  &+|\phi_{00}\rangle\langle\phi_{01}|\cdot\psi_{\sqrt{\gamma}\alpha}(p)\psi^{*}_{\sqrt{\gamma}\alpha e^{\frac{2\pi i}{3}} }(p)\\
  &+|\phi_{02}\rangle\langle\phi_{00}|\cdot \psi_{\sqrt{\gamma}\alpha e^{-\frac{2\pi i}{3}} }(p)\psi^{*}_{\sqrt{\gamma}\alpha}(p)\\
  &+|\phi_{02}\rangle\langle\phi_{01}|\cdot \psi_{\sqrt{\gamma}\alpha e^{-\frac{2\pi i}{3}} }(p)\psi^{*}_{\sqrt{\gamma}\alpha e^{\frac{2\pi i}{3}} }(p)\\
 &+|\phi_{01}\rangle\langle\phi_{00}|\cdot \psi_{\sqrt{\gamma}\alpha e^{\frac{2\pi i}{3}} }(p)\psi^{*}_{\sqrt{\gamma}\alpha}(p)\\
 &+\left.|\phi_{00}\rangle\langle\phi_{02}|\cdot \psi_{\sqrt{\gamma}\alpha e^{\frac{2\pi i}{3}} }(p)\psi^{*}_{\sqrt{\gamma}\alpha e^{-\frac{2\pi i}{3}} }(p)\right).
 \end{aligned}
\end{equation}
If we only accept the selection window $w_{0}=[-\Delta,\Delta]$, the
resulting unnormalized state is obtained by doing the $p$-integration,
\begin{equation}
 \sigma_{w_{0}}^{C_{0}}=\int\limits_{-\Delta}^{\Delta} dp~ \sigma_{p}^{C_{0}}.
\end{equation}
For carefully chosen $\Delta,\alpha$ and distance $L_{0}$, the contribution of the off-diagonal terms in
Eq.\eqref{eq: qutritp} can be neglected such that we obtain
the effective unnormalized state 
\begin{equation}
\label{eq: unnorm}
\begin{aligned}
 \widetilde{\rho}_{w_{0}}^{C_{0}}&=\frac{1}{3}\left(|\phi_{00}\rangle\langle\phi_{00}|\cdot \int\limits_{-\Delta}^{\Delta} dp~|\psi_{\sqrt{\gamma}\alpha}(p)|^{2}\right.\\
 &+|\phi_{02}\rangle\langle\phi_{02}|\cdot \int\limits_{-\Delta}^{\Delta} dp~|\psi_{\sqrt{\gamma}\alpha e^{-\frac{2\pi i}{3}} }(p)|^{2}\\
 &+\left.|\phi_{01}\rangle\langle\phi_{01}|\cdot \int\limits_{-\Delta}^{\Delta} dp~|\psi_{\sqrt{\gamma}\alpha e^{\frac{2\pi i}{3}} }(p)|^{2}\right).
\end{aligned}
 \end{equation}
The same calculation as above for $|C_{0}\rangle$ can be made for the other
two components in $\rho_{out}$ of Eq. \eqref{eq: qutritout}, $|C_{1}\rangle$ and $|C_{2}\rangle$.
The total conditional (unnormalized) density matrix then becomes 
\begin{equation}
 \begin{aligned}
   \widetilde{\rho}_{w_{0}}&=\frac{N_{u}(\sqrt{1-\gamma}\alpha)}{9}\widetilde{\rho}_{w_{0}}^{C_{0}}
   +\frac{N_{v}(\sqrt{1-\gamma}\alpha)}{9}\widetilde{\rho}_{w_{0}}^{C_{1}}\\
   &+\frac{N_{w}(\sqrt{1-\gamma}\alpha)}{9}\widetilde{\rho}_{w_{0}}^{C_{2}},
 \end{aligned}
\end{equation}
whose norm is the success probability,
\begin{equation}
\begin{aligned}
 p_{w_{0}}&=\text{Tr}[\widetilde{\rho}_{w_{0}}]\\
 &=\frac{1}{3}\int\limits_{-\Delta}^{\Delta} dp~\left(|\psi_{\sqrt{\gamma}\alpha}(p)|^{2}
 +|\psi_{\sqrt{\gamma}\alpha e^{-\frac{2\pi i}{3}} }(p)|^{2}\right.\\
 &+\left.|\psi_{\sqrt{\gamma}\alpha e^{\frac{2\pi i}{3}} }(p)|^{2}\right),
 \end{aligned}
\end{equation}
where we used $\text{Tr}[\rho_{out}]=1$
and $\text{Tr}[ \widetilde{\rho}_{w_{0}}^{C_{0}}]=\text{Tr}[ \widetilde{\rho}_{w_{0}}^{C_{1}}]=\text{Tr}[ \widetilde{\rho}_{w_{0}}^{C_{2}}]$.
The corresponding fidelity for the target state is then calculated as
\begin{equation}
 \begin{aligned}
&F_{w_{0}}=\frac{\langle \phi_{00}|\widetilde{\rho}_{w_{0}}|\phi_{00}\rangle}{ p_{w_{0}}}\\
&=\frac{N_{u}(\sqrt{1-\gamma}\alpha)}{9}~\frac{\frac{1}{3}\int\limits_{-\Delta}^{\Delta} dp~|\psi_{\sqrt{\gamma}\alpha}(p)|^{2}}{p_{w_{0}}}.
 \end{aligned}
\end{equation}
\noindent
The success probabilities for the other two selection windows are obtained in complete analogy,
\begin{equation}
 \begin{aligned}
   p_{w_{1}}&=\frac{1}{3}\int\limits_{\frac{\sqrt{3}}{2}\sqrt{\gamma}\alpha-\Delta}^{\infty} dp~\left(|\psi_{\sqrt{\gamma}\alpha}(p)|^{2}
 +|\psi_{\sqrt{\gamma}\alpha e^{-\frac{2\pi i}{3}} }(p)|^{2}\right.\\
 &+\left.|\psi_{\sqrt{\gamma}\alpha e^{\frac{2\pi i}{3}} }(p)|^{2}\right),\\
  p_{w_{2}}&=\frac{1}{3}\int\limits_{-\infty}^{-\frac{\sqrt{3}}{2}\sqrt{\gamma}\alpha+\Delta} dp~\left(|\psi_{\sqrt{\gamma}\alpha}(p)|^{2}
 +|\psi_{\sqrt{\gamma}\alpha e^{-\frac{2\pi i}{3}} }(p)|^{2}\right.\\
 &+\left.|\psi_{\sqrt{\gamma}\alpha e^{\frac{2\pi i}{3}} }(p)|^{2}\right).\\
 \end{aligned}
\end{equation}
\noindent
The corresponding fidelities with respect to the target states $|\phi_{01}\rangle$ and $|\phi_{02}\rangle$
for these windows are, respectively,
\begin{equation}
 \begin{aligned}
F_{w_{1}}&=\frac{N_{v}(\sqrt{1-\gamma}\alpha)}{9}~\frac{\frac{1}{3}\int\limits_{\frac{\sqrt{3}}{2}\sqrt{\gamma}\alpha-\Delta}^{\infty}~|\psi_{\sqrt{\gamma}\alpha}(p)|^{2}}{
p_{w_{1}}},
 \end{aligned}
\end{equation}
and 
\begin{equation}
 \begin{aligned}
F_{w_{2}}&=\frac{N_{w}(\sqrt{1-\gamma}\alpha)}{9}~\frac{\frac{1}{3}\int\limits_{-\infty}^{-\frac{\sqrt{3}}{2}\sqrt{\gamma}\alpha+\Delta}~|\psi_{\sqrt{\gamma}\alpha}(p)|^{2}}{
 p_{w_{2}}}.
 \end{aligned}
\end{equation}
To estimate the performance of this entanglement generation scheme, 
we define the average fidelity as
\begin{equation}
 F_{av}=\frac{\sum\limits_{i=0}^{2}p_{w_{i}}F_{w_{i}}}{P_{succ}},
\end{equation}
where $P_{succ}=\sum\limits_{i=0}^{2}p_{w_{i}}$ is the total success probability.
The $\alpha$-dependence of the success probability and the average fidelity for various values of $\Delta$
is shown in Figs. \ref{fig: qutritprob} and \ref{fig: qutritfid} for $L_{0}$=5~km.\\
Clearly, if $\Delta=\Delta_{0}$, then there is no failure window at all and all measurement results
are accepted. This corresponds to unit success probability, $P_{succ}=1$. On the other hand, for smaller (but not too small)
$\Delta$, i.e., $\Delta<\Delta_{0}$,
the success probability still tends to unity for increasing $\alpha$, as long as the three coherent states remain well within
their respective selection windows.
The fidelity, however, shows an opposite behavior. The smaller $\Delta$ is chosen, the higher 
the average fidelity for moderate values of $\alpha$. Increasing $\alpha$ makes the fidelity finally drop to 1/3, which is a direct
consequence of the loss channel whose mixed output becomes more and more balanced for larger $\alpha$. For each
chosen value of $\Delta$, there is an optimal value for $\alpha$ leading to a maximal fidelity.
For instance, still with  $L_{0}$=5~km, choosing $\Delta=0.2\Delta_{0}$ and $\alpha\approx 1$ leads to an average fidelity of $F_{av}\approx 0.7$
at a very reasonable success probability of $P_{succ}\approx 0.4$. The corresponding plots for elementary distances of  $L_{0}$=10~km
are shown in Figs. \ref{fig: qutritprob2} and \ref{fig: qutritfid2}.
A possible ququart scheme for distributing ququart-ququart entanglement is explicitly discussed in App. \ref{sec:ququart}.

\subsection{Unambiguous state discrimination}
\label{sec: USD}
\begin{figure}[t!]
\centering
\includegraphics[width=0.5\textwidth]{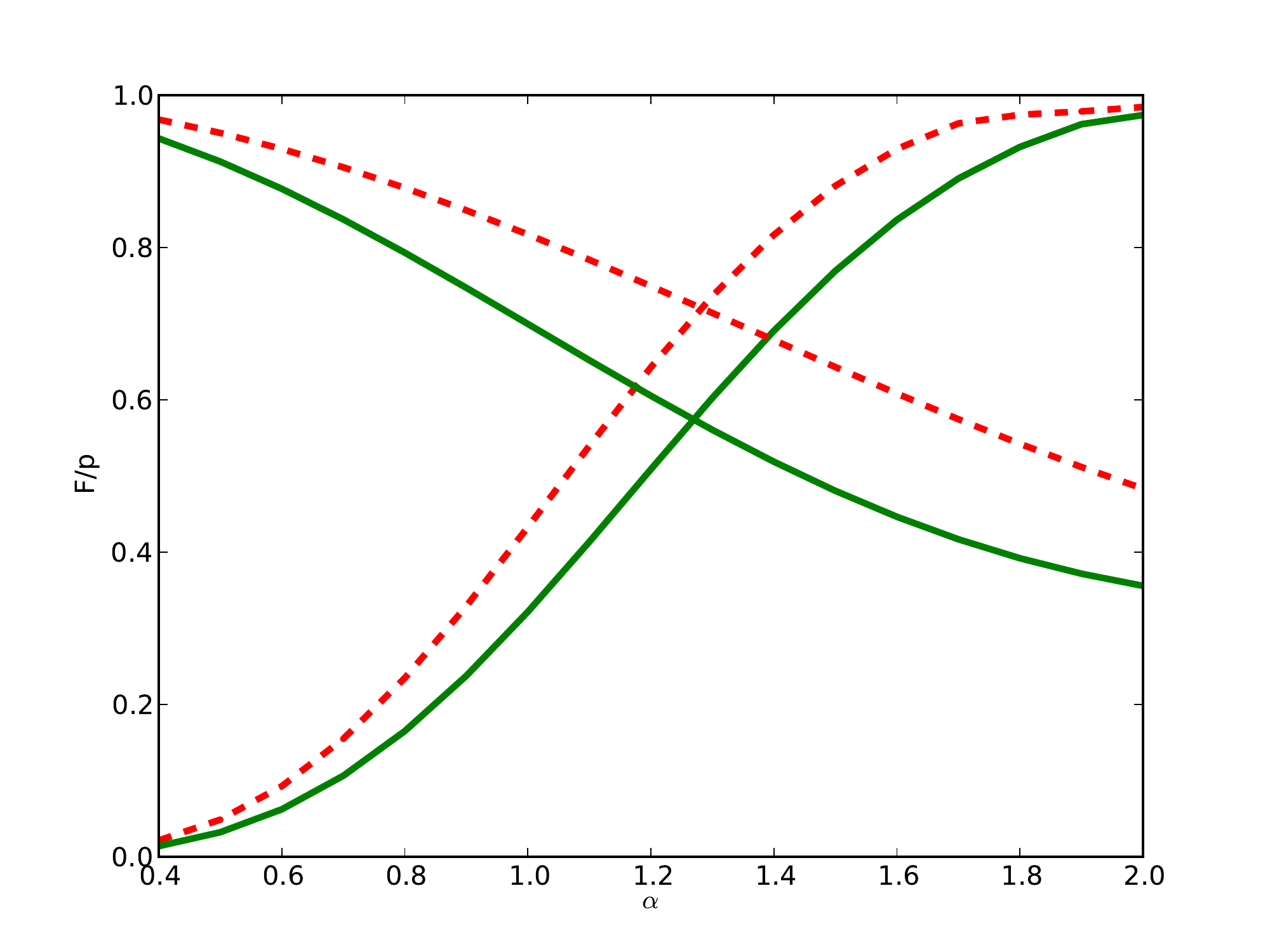}
\caption{Success probabilities and fidelities for the USD-based scheme for 
5 km (red, dotted) and 10 km (green, solid) in dependence of $\alpha$.
}
\label{fig: discrimination}
\end{figure}

In this section, we will consider an alternative measurement scheme for a qutrit hybrid repeater based 
upon so-called unambiguous state discrimination (USD). Compared to the homodyne-based scheme,
the conceptual difference in the USD-based scheme is that the non-orthogonality of the coherent states
only affects $P_{succ}$ and no longer $F_{av}$, as USD enables one to discriminate
non-orthogonal states probabilistically in an error-free fashion. 
The idea is that a successful and error-free projection onto one of the states 
$|\sqrt{\gamma}\alpha\rangle$ or $|\sqrt{\gamma}\alpha e^{\pm\frac{2\pi i}{3}}\rangle$
would lead to maximally entangled states in all components in Eq.\eqref{eq: qutritout}.
The task  is therefore to find the most efficient possible scheme in the framework
of quantum theory for unambiguously discriminating between the three coherent states
above.\\
This problem was treated by Chefles \cite{Chefles} 
who derived the optimal success probability as
\begin{equation}
\label{eq: qutritUSD}
 P_{D}\leq \min\limits_{r} \sum\limits_{j=0}^{2}e^{-\frac{2\pi i j r}{3}}e^{\gamma\alpha^{2}(e^{\frac{2\pi i j}{3}}-1)},
\end{equation}
with $r=0,1,2$ (see also Refs. \cite{vanEnk, Croke}). The relation between this optimal probability and the corresponding fidelity
of the final maximally entangled state is shown in Fig. \ref{fig: discrimination}.\\ 

 \subsection{Entanglement purification}
After the homodyne detection, the conditional state state resulting from  Eq. \eqref{eq: qutritout}
still represents a mixed state. Depending on the channel distance,
the selection window, and the amplitude $\alpha$, the 
resulting state in the first component is 
a mixture of the dominant target state $|\phi_{00}\rangle$
with small extra components of $|\phi_{02}\rangle$ and $|\phi_{01}\rangle$ (if the result belongs to window $w_{0}$).
This is similar for the other two components of the mixture with their rotated Bell states.
Thus, effectively, the state after homodyne detection reads as
(up to local qutrit rotations in case of the other two windows)
\begin{equation}
\label{eq: qutriteff}
\begin{aligned}
  \rho_{eff}&=\frac{N_{u}(\sqrt{1-\gamma}\alpha)}{9}|\widetilde{C}_{0}\rangle\langle \widetilde{C}_{0}|\\
  &+\frac{N_{v}(\sqrt{1-\gamma}\alpha)}{9}|\widetilde{C}_{1}\rangle\langle \widetilde{C}_{1}|\\
  &+\frac{N_{w}(\sqrt{1-\gamma}\alpha)}{9}|\widetilde{C}_{2}\rangle\langle \widetilde{C}_{2}|,
 \end{aligned}
\end{equation}
where 
\begin{equation}
\label{eq: components}
 \begin{aligned}
  |\widetilde{C}_{0}\rangle&=\frac{1}{\sqrt{3}}(|00\rangle+|11\rangle+|22\rangle),\\
  |\widetilde{C}_{1}\rangle&=\frac{1}{\sqrt{3}}(|00\rangle+e^{-\frac{2\pi i}{3}}|11\rangle+e^{\frac{2\pi i}{3}}|22\rangle),\\
  |\widetilde{C}_{2}\rangle&=\frac{1}{\sqrt{3}}(|00\rangle+e^{+\frac{2\pi i}{3}}|11\rangle+e^{-\frac{2\pi i}{3}}|22\rangle).
 \end{aligned}
\end{equation}
Note that in the case of USD, Eqs. \eqref{eq: qutriteff} and \eqref{eq: components}  represents the exact
output state
and there are no extra terms from the rotated Bell states
(which nonetheless can be neglected for the case of homodyne detection provided
the selection window-based state discrimination works sufficiently well).
In general, mixed entangled states degrade the performance
of quantum information processing tasks like teleportation
or the entanglement swapping operation discussed in the next section. Hence,
a purification of the above mixed state is required.\\
Entanglement purification aims at generating fewer high-fidelity
copies from many noisy copies of a certain pure target state via local operations and classical communication.
By iterating this purification
protocol, a fidelity  arbitrarily close to unity  can be achieved.
The purification of mixed qubit states was investigated by Bennett et. al \cite{Bennett}
for the class of Werner states \cite{Werner2}. Nearly at the same time, Deutsch
et al. \cite{Deutsch} demonstrated a similar purification
protocol for states diagonal in the Bell basis. This protocol requires
only two copies for each step and leads to a better efficiency compared to the Bennett scheme.
The latter was demonstrated experimentally \cite{Pan2001,Pan2003} and also generalized to arbitrary dimensions \cite{Horodecki,Alber}.\\
To perform a purification of our relevant state, i.e. to increase the statistical weight of $|\widetilde{C}_{0}\rangle$
in Eq. \eqref{eq: qutriteff}, at least two copies of the matter-matter output state are required.
On each copy, the following transformations are performed:
The first matter  qutrit system is subject to the transformation
\begin{equation}
\begin{aligned}
 |0\rangle&\mapsto \frac{1}{\sqrt{3}}(|0\rangle+|1\rangle+|2\rangle),\\
 |1\rangle&\mapsto \frac{1}{\sqrt{3}}(|0\rangle+e^{i\phi}|1\rangle+e^{-i\phi}|2\rangle),\\
 |2\rangle&\mapsto \frac{1}{\sqrt{3}}(|0\rangle+e^{-i\phi}|1\rangle+e^{i\phi}|2\rangle),\\
\end{aligned}
\end{equation}
while on the second system,
\begin{equation}
\begin{aligned}
 |0\rangle&\mapsto \frac{1}{\sqrt{3}}(|0\rangle+|1\rangle+|2\rangle),\\
 |1\rangle&\mapsto \frac{1}{\sqrt{3}}(|0\rangle+e^{-i\phi}|1\rangle+e^{i\phi}|2\rangle),\\
 |2\rangle&\mapsto \frac{1}{\sqrt{3}}(|0\rangle+e^{i\phi}|1\rangle+e^{-i\phi}|2\rangle),\\
\end{aligned}
\end{equation}
is performed where $\phi=\frac{2\pi}{3}$. The components of the mixture are then transformed
as
\begin{equation}
 \begin{aligned}
  |\widetilde{C}_{0}\rangle&\mapsto \frac{1}{\sqrt{3}}(|00\rangle+|11\rangle+|22\rangle),\\
  |\widetilde{C}_{1}\rangle&\mapsto\frac{1}{\sqrt{3}}(|01\rangle+|12\rangle+|20\rangle),\\
  |\widetilde{C}_{2}\rangle&\mapsto  \frac{1}{\sqrt{3}}(|10\rangle+|21\rangle+|02\rangle).
 \end{aligned}
\end{equation}
\noindent 
A mixture of $|\widetilde{C}_{0}\rangle$, $|\widetilde{C}_{1}\rangle$,
and $|\widetilde{C}_{2}\rangle$
with statistical weights $p_{0}$, $p_{1}$, and $p_{2}$, where $p_{0}+p_{1}+p_{2}=1$,
can now be purified as follows.
One takes two copies of the state that is shared between two parties $A$ and $B$.
As proven in Sec. \ref{sec: qudit} for arbitrary dimensions, 
local subtraction gates are applied on the qutrits belonging
to A and B. After this, A and B select one of the two copies and measure its respective spin.
Equal spin results lead to the new mixed state
\begin{equation}
 \rho^{\prime}=\frac{\sum\limits_{j=0}^{2}p_{j}^{2}|\widetilde{C}_{j}\rangle\langle \widetilde{C}_{j}|}{\sum\limits_{j=0}^{2}p_{j}^{2}},
\end{equation}
whose fidelity with respect to the target state
$|\widetilde{C_{0}}\rangle$ is now increased, provided $p_{0}>1/3$ and $p_{1}, p_{2}<p_{0}$.
\subsection{Entanglement swapping}
In the previous sections, we have shown how to entangle two qutrits over a distance $L_{0}$.
The distance $L_{0}$, however, is typically to short for general applications in quantum communication.
It is therefore necessary to further extend the entanglement over larger distances. This can be done 
by entanglement swapping.\\
To perform entanglement swapping, two entangled qutrit-qutrit pairs are generated next to each other,
covering a total distance of $2L_{0}$. To connect the two pairs
and thus distribute entanglement over twice the initial distance, a Bell measurement is 
carried out on the two adjacent matter systems. A successful Bell measurement projects the remaining
two matter systems onto a maximally entangled state.\\ 
In analogy to the qubit case, a Bell measurement on two qutrits can be performed by 
applying a qudit sum gate (\textsc{cnot} or \textsc{cshift}), followed by measurements in the $X$ and in the $Z$ basis (see Eq. \eqref{eq: qutritz}).
As pointed out in \cite{Dusek}, Hadamard transformations and 
a \textsc{cphase} gate suffice to implement the sum gate.\\
In the following, we assume that arbitrary single qutrit rotations and measurements can be performed 
on the matter systems and show how to construct the sum gate based on these assumptions.\\
In our framework, a \textsc{cphase} gate is represented by the unitary operation
\begin{equation}
 U_{CP}=\exp\left(-\frac{2\pi i}{3}S_{z_{1}}^{(3)}S_{z_{2}}^{(3)}\right),
\end{equation}
where the operators $S_{z_{i}}^{(3)}$ correspond to the operations introduced in Eq.\eqref{eq: qutritz} on the ith qutrit. 
Like in the qubit case of a \textsc{cnot} gate, a decomposition for the qudit \textsc{cshift} gate is given by
\begin{equation}
 \operatorname{\textsc{cshift}}=(H\otimes \mathbbm{1})\cdot\operatorname{\textsc{cphase}}\cdot(H\otimes \mathbbm{1}),
\end{equation}
where $H$ is the qutrit Hadamard transformation. Indeed,
one observes by direct calculation
$(H\otimes \mathbbm{1})\operatorname{\textsc{cphase}}(H\otimes \mathbbm{1})|x,y\rangle=|x\ominus y,y\rangle$
for $x,y\in \mathbb{Z}_{2}$. Note that $\ominus$
denotes subtraction modulo 3. A more formal proof of this decomposition
for arbitrary dimensions is given in Sec. \ref{sec: qudit}.\\
With HQR protocols for qubits and qutrits in mind,
an extension to ququarts, i.e. four-level systems, is straightforward. As a bridge to
the general qudit case, as presented in the next section, it is nonetheless
useful to also explicitly consider the ququart case including the optical qubus measurements adapted to this case.
It is presented in App. \ref{sec: rate}.

\subsection{Rate analysis}
\label{sec: rate2}
\subsubsection{Methods and assumptions}
In this section, we quantify the performance of our qutrit HQR protocol for the generation of entanglement over
the total channel distance $L_{0}$. The performance can be assessed by the entanglement generation rate, i.e., the number of entangled pairs
over the entire distance per unit time. Besides this, the fidelity of the generated states is of particular interest.\\
The atomic matter systems also serve as quantum memories (as needed because of the probabilistic step of entanglement
purification after the entanglement distribution) and we assume matter systems with infinite coherence time, i.e. perfect memories.
In addition, we assume deterministic and error-free gates on them. Especially,
the entanglement swapping operation is treated as deterministic employing the gates as described in the preceeding section. Strictly speaking, photon transmission loss
is the only error source entering our rate analysis
and the resulting rates have to be understood as upper bounds of the actual achievable rates. For this scenario, analytical formulae 
for the rates in dependence of the number of elementary segments as well as the number of purifications performed on each segment
after the distributions have been derived in \cite{rate}. Note that we include one to several rounds of entanglement purification only right after
the initial entangled-state distributions. In this theoretical treatment, our repeater scheme effectively becomes a second
generation quantum repeater (recall Sec. \ref{sec: intro}) where rates are ultimately limited by $R\lesssim\frac{c}{L_{0}}$
(instead of $R\lesssim \frac{c}{L}$ if purifications were performed until the final nesting level \cite{Briegel1,Briegel2}) \cite{Ultrafast, Optiarchi}.
\\
We consider $2^{n}$ segments of elementary distance $L_{0}$, covering a total distance $L=2^{n}L_{0}$. Entanglement is generated 
in each segment with a probability $P_{0}$. If the obtained state is not directly purified, the resulting rate becomes
\begin{equation}
 R_{n}=\frac{c}{2L_{0}}\frac{1}{Z_{n}(P_{0})}=\frac{1}{T_{0}Z_{n}(P_{0})}
\end{equation}
where
\begin{equation}
 Z_{n}(P_{0})=\sum\limits_{j=1}^{2^{n}}\binom{2^{n}}{j}\frac{1}{1-(1-P_{0})^{j}}
\end{equation}
is the average total number of attempts it takes for all segments to eventually
share an entangled pair (recall that initially shared pairs can be stored as long as needed),
$T_{0}=\frac{2L_{0}}{c}$ is the elementary time unit for sending the quantum states and also the classical information
to confirm their successful distribution (as well as purification), 
and $c$ is the speed of light in the optical fiber.\\
If one round of purification is performed, the same formula can be applied, but now $P_{0}$
has to be substituted by an effective probability,
\begin{equation}
 Q_{1}(L_{0})=P_{0}P_{1}\left(\frac{2-P_{0}}{3-2P_{0}}\right),
\end{equation}
where $P_{1}$ is the probability for the first round of purification to succeed. Furthermore,
the rates with two and three rounds of purification can be calculated using the effective probabilities
\begin{equation}
 Q_{2}(L_{0})= Q_{1}(L_{0})P_{2}\left(\frac{2- Q_{1}(L_{0})}{3-2 Q_{1}(L_{0})}\right),
\end{equation}
and
\begin{equation}
 Q_{3}(L_{0})= Q_{2}(L_{0})P_{3}\left(\frac{2- Q_{2}(L_{0})}{3-2 Q_{2}(L_{0})}\right),
\end{equation}
where $P_{2}$ and $P_{3}$ are the success probabilities for two and three rounds of purification, respectively.
Note that without the use of quantum memories, $Q_{3}$ would scale as $P_{0}^{8}P_{1}^{4}P_{2}^{2}P_{3}$,
which (assuming small probabilities) is turned into a scaling like $P_{0}P_{1}P_{2}P_{3}$ 
with the help of the quantum memories.
Higher rounds of purification can be considered in a recursive fashion. We analyze the rates for the USD- and homodyne-based scheme
separately in the next two sections.
\subsubsection{USD-based scheme}
For the USD-scheme, $P_{0}$ is given by the optimal probability in Eq.\eqref{eq: qutritUSD} to distinguish the three coherent states
$|\sqrt{\gamma}\alpha\rangle$ and  $|\sqrt{\gamma}\alpha e^{\frac{2\pi i}{3}}\rangle$. The resulting state is the normalized version of Eq. \eqref{eq: unnorm} and the initial fidelity of 
the target state reads as
\begin{equation}
 F_{0}=\frac{N_{u}(\sqrt{1-\gamma}\alpha)}{9},
\end{equation}
and 
\begin{equation}
 \begin{aligned}
  F_{1}&=\frac{N_{v}(\sqrt{1-\gamma}\alpha)}{9},\\
  F_{2}&=\frac{N_{w}(\sqrt{1-\gamma}\alpha)}{9},
 \end{aligned}
\end{equation}
for the other two components. One round of purification succeeds with 
probability
\begin{equation}
 P_{1}=F_{0}^{2}+F_{1}^{2}+F_{2}^{2},
\end{equation}
and the resulting improved fidelity is
\begin{equation}
 F_{0}^{\prime}=\frac{F_{0}^{2}}{F_{0}^{2}+F_{1}^{2}+F_{2}^{2}}.
\end{equation}
For more rounds of purification, the fidelities and success probabilities can be obtained recursively.\\
After entanglement swapping, the final fidelity of the entangled state distributed over the total
distance is lower bounded by $(\widetilde{F}_{0})^{2^{n}}$, where $\widetilde{F}_{0}$ is the final fidelity
for each segment, possibly obtained after some rounds of purification.

\subsubsection{Homodyne-based scheme}
An exact rate analysis for the HQR with entanglement distribution based 
on homodyne detection is much more demanding than for the USD-case. This is due to the fact
that at adjacent elementary segment potentially different mixed quantum states are generated depending
on the corresponding measurement result. As already pointed out, these states can be brought into a similar 
form, i.e., the components are equal, but the statistical weights are not necessarily equal. An exact rate analysis
is therefore out of reach.\\
To nevertheless assess the performance of that scheme, we model the situation with an effective state on each elementary segment. This effective
state has the average fidelity $F_{av}(\alpha,\gamma)$ as the statistical weight of the first component, whereas the other two components 
are equally weighted with $F_{1}=F_{2}=\frac{1}{2}(1-F_{av}(\alpha,\gamma))$. For an elementary distance of $L_{0}=5$ km, we choose
$\alpha\approx 1$, which leads to a maximum initial fidelity of $\approx 0.7$. As the generation probability $P_{0}$, we insert the success probability, $P_{succ}=\sum\limits_{i=0}^{2}p_{w_{i}}$, for obtaining
a result in one of the success windows (see Sec. \ref{sec: matter-matter}) which equals $\approx 0.4$ in this case. For $L_{0}=10$ km, we also
have $\alpha \approx 1$, but now $F_{av}\approx 0.6$ and $P_{0}\approx 0.39$.\\
Using these initial values, the formulae for the rates and fidelities, including some possible rounds of purification, can directly be applied.
For quantitative examples and an illustration of the trade-off between repeater rates and fidelities, see App. \ref{sec: rate}.\\
To summarize some of the results presented there, for elementary distances as short as $L_{0}=5$ km, the USD-based scheme and the homodyne-based
scheme perform comparably. In either case at least three rounds of purification are needed in order to obtain reasonable fidelities
and rates for distances as large as 640 km. 
For $L_{0}=10$ km according to our calculations, the USD-based scheme performs slightly better than the homodyne-based scheme, such that in both scenarios
rather high fidelities can be achieved for distances as large as 1280 km (the rates are comparable and again three rounds of purification are necessary). 
However, note that our results for the homodyne-based scheme only hold under the assumptions that the off-diagonal terms in Eq.\eqref{eq: qutritp}
are negligible and that the conditional state after homodyne detection can be modeled via an effective state with fidelity $F_{av}$.
Thus, the numbers presented in App. \ref{sec: rate} may overestimate the homodyne-based scheme compared to the USD-based scheme.\\
Results for a situation with a more practical repeater spacing, $L_{0}=20$ km, indicate that for $L=1280$ km near-unit fidelities at rates
$\sim$ Hz are only achievable using USD, because in the homodyne-based scheme the output fidelities drop below 0.5 for such large elementary
distances. Note that a similar observation was made for the original qubit scheme based on homodyne detection \cite{Bright}.

\section{The general qudit case}
\label{sec: qudit}
\begin{figure}[t!]
\centering
\includegraphics[width=0.35\textwidth]{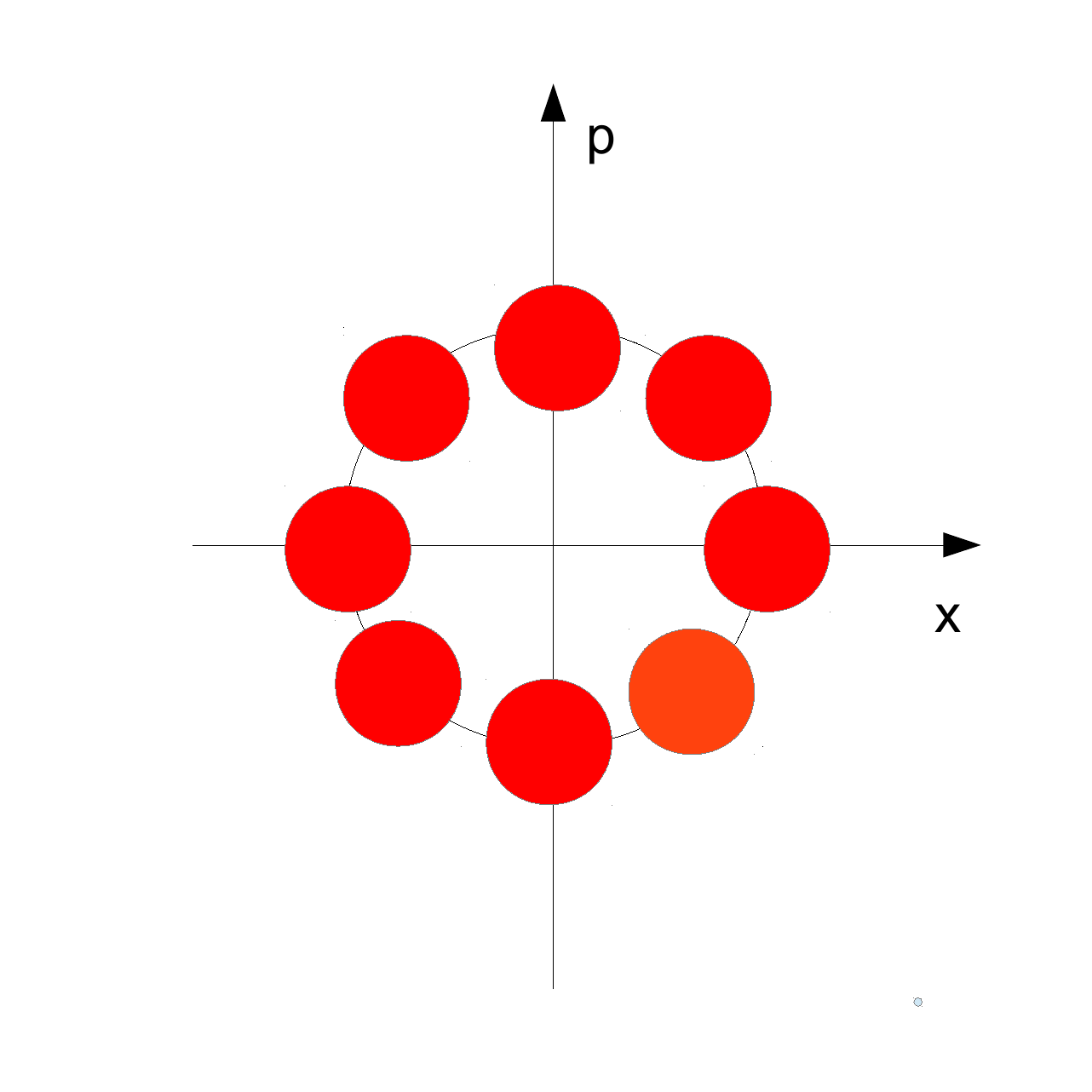}
\caption{Phase space representation of the qubus mode for $d=8$.}
\label{fig: generalpic}
\end{figure}

Based on the results obtained in the last sections for specific examples,
we are now in turn to propose HQR protocols for arbitrary 
finite dimensional quantum systems.\\
The dispersive interaction between a general qudit, i.e. a $d$-level system,
and a light mode can be realized by the Hamiltonian
\begin{equation}
 H_{int}^{(d)}=\hbar g S_{z}^{(d)} a^{\dagger}a
\end{equation}
with $S_{z}^{(d)}|k\rangle=\left(\frac{2k-d+1}{2}\right)|k\rangle$ for $k=\{0,1,..,d-1\}$,
and where $S_{z}^{(2)}=\sigma_{z}$. The corresponding unitary
is $U_{d}(\theta)=\exp(i\theta S_{z}^{(d)} a^{\dagger}a )$ and the relevant case
of a strong interaction is obtained by setting $\theta=\frac{2\pi}{d}$.\\
The first step in the protocol is the preparation of
the matter state $\frac{1}{\sqrt{d}}\sum\limits_{k=0}^{d-1}|k\rangle$, which then
interacts with an optical coherent state $|\alpha\rangle$ via the strong dispersive interaction. 
This results in a hybrid entangled qudit-light (qudit-qubus) state,
\begin{equation}
 \frac{1}{\sqrt{d}}\sum\limits_{k=0}^{d-1}|k\rangle|\alpha e^{\frac{2k\pi ik }{d}}\rangle.
\end{equation}
After locally  generating qudit-light entanglement, the light mode
is sent through an optical channel of length $L_{0}$ where it
is subject to photon loss.
Including again an ancilla vacuum mode and mixing it with the optical mode
results in 
\begin{equation}
\label{eq: beforetrace}
 \frac{1}{\sqrt{d}}\sum\limits_{q=0}^{d-1}|q\rangle|\sqrt{\gamma}\alpha e^{\frac{2\pi i q}{d}}\rangle|\sqrt{1-\gamma}\alpha e^{\frac{2\pi i q}{d}}\rangle.
\end{equation}
\noindent
As in the specific examples above, the crucial point is now to find a suitable basis for tracing out the loss mode.
Here, in the general case, this basis consists of the $d$ vectors
\begin{equation}
\label{eq: quditbasis}
 |v_{m}\rangle=\frac{1}{\sqrt{N_{v_{m}}(\alpha)}}\sum\limits_{k=0}^{d-1}e^{\frac{2\pi i km}{d}}|\alpha e^{\frac{2\pi i k}{d}}\rangle,
\end{equation}
with $m=0,1,..,d-1$. We can thus recast the coherent states of the ancilla light mode
in Eq. \eqref{eq: beforetrace} as
\begin{equation}
 |\alpha e^{\frac{2\pi i k}{d}}\rangle=\frac{1}{d}\sum\limits_{m=0}^{d-1}\sqrt{N_{v_{m}}(\alpha)}e^{-\frac{2\pi i km}{d}}|v_{m}\rangle,
\end{equation}
and find for Eq. \eqref{eq: beforetrace}:
\begin{equation}
 \frac{1}{d\sqrt{d}}\sum\limits_{q,m=0}^{d-1}\sqrt{N_{v_{m}}(\sqrt{1-\gamma}\alpha)}e^{-\frac{2\pi i qm}{d}}|q\rangle|\sqrt{\gamma}\alpha e^{\frac{2\pi i q}{d}}\rangle|v_{m}\rangle.
\end{equation}
Tracing out the loss mode in this basis is now a trivial task and one obtains
\begin{equation}
\label{eq: quditout}
 \begin{aligned}
\rho_{out}&=\sum\limits_{m=0}^{d-1}\frac{N_{v_{m}}(\sqrt{1-\gamma}\alpha)}{d^{2}}\\
&\times\left[\left(\frac{1}{\sqrt{d}}\sum\limits_{q=0}^{d-1}e^{-\frac{2\pi i qm}{d}}|q\rangle|\sqrt{\gamma}\alpha e^{\frac{2\pi i q}{d}}\rangle\right)
\times H.c.\right]
 \end{aligned}
\end{equation}
for the $d$-component qudit-light output state.\\
Again, this can be further simplified by basis transformations on both the light mode and the matter system.
The light mode can be expressed in the basis given in Eq. \eqref{eq: quditbasis}, while the matter system can be written in the 
(generalized Pauli) qudit \textit{X}-basis,
\begin{equation}
 |\widetilde{k}\rangle=\frac{1}{\sqrt{d}}\sum\limits_{m=0}^{d-1}e^{\frac{2\pi i km}{d}}|m\rangle,
\end{equation}
for $k=0,1,...,d-1$. This gives the expression
\begin{equation}
 \begin{aligned}
\rho_{out}&=\sum\limits_{m=0}^{d-1}\frac{N_{v_{m}}(\sqrt{1-\gamma}\alpha)}{d^{2}}\\
\times&\left[\left(\frac{1}{d}\sum\limits_{r=0}^{d-1}\sqrt{N_{v_{r}}}|\widetilde{m\oplus r}\rangle|\widetilde{v}_{r}\rangle\right)
\times H.c.\right]
 \end{aligned}
\end{equation}
for Eq. \eqref{eq: quditout} where $\oplus$
denotes addition modulo $d$. Note that $\sim$
again indicates basis vectors with damped amplitude $\sqrt{\gamma}\alpha$ on the light mode
and the $\textit{X}$-basis on the matter system.\\
After traveling through the loss channel over a distance $L_{0}$, 
the light mode reaches a second 
matter system, also prepared in the state $\frac{1}{\sqrt{d}}\sum\limits_{k=0}^{d-1}|k\rangle$.
The light mode interacts dispersively with the second matter system, this time with the inverse
angle $\theta=-\frac{2\pi}{d}$. The resulting state becomes
\begin{equation}
 \rho=\sum\limits_{m=0}^{d-1}\frac{N_{v_{m}}}{d^{2}} |T_{m}\rangle\langle T_{m}|,
\end{equation}
with the components
\begin{equation}
 |T_{m}\rangle=\frac{1}{d}\sum\limits_{q=0}^{d-1}\sum\limits_{l=0}^{d-1}e^{-\frac{2\pi i qm}{d}}|q\rangle|l\rangle|\sqrt{\gamma}\alpha e^{\frac{2\pi i (q-l)}{d}}\rangle,
\end{equation}
written in the original basis (like in Eq.\eqref{eq: quditout}).\\
The state discrimination in the general case involves 
the $d$ coherent states $|\sqrt{\gamma}\alpha\rangle,...,|\sqrt{\gamma}\alpha e^{\frac{2\pi i (d-1)}{d}}\rangle$
which can be graphically represented as coherent states "on a ring" (see Fig. \ref{fig: generalpic}~ for $d=8$). 
A projection onto one of the $d$ coherent states collapses each component
onto a maximally entangled state. However, by increasing the dimension $d$,
a projection scheme based on homodyne detection becomes more and more futile since 
no direction is uniquely specified any more.
\\
A scheme for unambiguously discriminating exactly these $d$ coherent states was derived in 
\cite{Chefles} for arbitrary dimensions (for $d=3$, recall Sec. \ref{sec: USD}). An upper bound for the success 
probability is given by
\begin{equation}
 P_{D}\leq \min\limits_{r} \sum\limits_{j=0}^{d-1}e^{-\frac{2\pi i j r}{d}}e^{\gamma\alpha^{2}(e^{\frac{2\pi i j}{d}}-1)},
\end{equation}
$r=0,1,..,d-1$, where Eq.\eqref{eq: qutritUSD} is recovered for $d=3$. Since the upper bound on the
right-hand side depends on both $\alpha$ and $\gamma$ the minimization 
with respect to $r$ is hard analytically. We therefore calculate the bound numerically.\\
After the USD, the resulting mixed state will be a mixture of $d-1$ maximally entangled Bell states
of the form
\begin{equation}
 |\phi_{kj}\rangle=\frac{1}{\sqrt{d}}\sum\limits_{y=0}^{d-1}e^{\frac{2\pi i k y}{d}}|y,y\ominus j\rangle,
\end{equation}
for one fixed $j=0,...,d-1$, according to the specific identified coherent state.\\
If $j\neq 0$, a $j$-fold application of $X=\sum\limits_{k=0}^{d-1}|k+1\rangle\langle k|$
transforms all these states to
\begin{equation}
 |\phi_{k0}\rangle=\frac{1}{\sqrt{d}}\sum\limits_{y=0}^{d-1}e^{\frac{2\pi i k y}{d}}|y,y\rangle.
\end{equation}
By means of local unitaries, the different components of the mixtures
with  $|\phi_{k0}\rangle$ can always be transformed to a mixture
of the states
\begin{equation}
 |\psi_{j}\rangle\equiv |\phi_{0j}\rangle= \frac{1}{\sqrt{d}}\sum\limits_{y=0}^{d-1}|y,y\ominus j\rangle,
\end{equation}
with now all $j$ included.
We therefore obtain 
\begin{equation}
\label{eq: topurify}
 \rho=\sum\limits_{j=0}^{d-1}p_{j}|\psi_{j}\rangle\langle\psi_{j}|
\end{equation}
for the state to be purified.\\
The purification now works as follows. We prepare two copies of the state
in Eq.\eqref{eq: topurify} such that the total joint four-qudit state reads
\begin{equation}
 \rho\otimes \rho=\sum\limits_{j=0}^{d-1}\sum\limits_{k=0}^{d-1}p_{j}p_{k}|\psi_{j}\rangle|\psi_{k}\rangle\langle\psi_{j}|\langle\psi_{k}|,
\end{equation}
where the individual terms are
\begin{equation}
 |\psi_{j}\rangle |\psi_{k}\rangle=\frac{1}{d}\sum\limits_{y=0}^{d-1}\sum\limits_{y^{\prime}=0}^{d-1}|y,y\ominus j\rangle|y^{\prime},y^{\prime}\ominus k\rangle.
\end{equation}
One applies a local CSHIFT gate on systems 1 and 3
as well 2 and 4 in order to obtain
\begin{equation}
\frac{1}{d}\sum\limits_{y=0}^{d-1}\sum\limits_{y^{\prime}=0}^{d-1}|y-y^{\prime},y\ominus y^{\prime}\oplus k\ominus j\rangle|y^{\prime},y^{\prime}\ominus k\rangle.
\end{equation}
After that, the first spins of the first two systems are measured.
If the spins are parallel, it follows $k=j$
such that only diagonal parts contribute. 
As a consequence, the second two systems collapse to $|\psi_{k}\rangle$.\\
The new state then becomes
\begin{equation}
 \rho^{\prime}=\frac{\sum\limits_{j=0}^{d-1}p_{j}^{2}|\psi_{j}\rangle\langle \psi_{j}|}{\sum\limits_{j=0}^{d-1}p_{j}^{2}}.
\end{equation}
The fidelity with respect to the target state $|\psi_{0}\rangle$
is thus
\begin{equation}
 F^{\prime}=\frac{p_{0}^{2}}{\sum\limits_{j=0}^{d-1}p_{j}^{2}},
\end{equation}
which is increased compared to the initial fidelity $p_{0}$ if $p_{0}>\frac{1}{d}$ and $p_{i}<p_{0}$ for $i=1,...,d-1$.\\
After possibly several rounds of purification, a high-fidelity entangled state
can be obtained between the two separated qudits. This is referred to as the initial entanglement generation
or distribution.\\
To further extend the  entanglement, two elementary segments next to each other are connected
via entanglement swapping through Bell measurements  on adjacent repeater nodes, i.e., a projection on maximally
entangled qudit-qudit states.\\
Generalizing the qutrit case, we show that 
the \textsc{cshift} gate lies at the heart of such Bell measurements
and that these be realized by a \textsc{cphase} gate based on 
the generalized dispersive interaction.
\\
The \textsc{cphase} gate for an arbitrary dimension $d$ is realized by the two-qudit unitary transformation
\begin{equation}
 U_{d}=\exp\left(-\frac{2\pi i}{d}S_{z_{1}}^{(d)}S_{z_{2}}^{(d)}\right),
\end{equation}
with the generalized spin operator $S_{i}^{(d)}$ acting on qudit $i$. We show
by direct calculation that the sequence $H\otimes 1\rightarrow \textsc{cphase}\rightarrow H\otimes 1$
acts as a controlled phase shift gate on an arbitrary two-qudit state:
\begin{equation}
 \begin{aligned}
&(H\otimes 1)\cdot \textsc{cphase}\cdot(H\otimes 1)|xy\rangle\\
&=(H\otimes 1)\cdot \textsc{cphase}\frac{1}{\sqrt{d}}\sum\limits_{k=0}^{d-1}\exp\left(\frac{2\pi i k x}{d}\right)|ky\rangle\\
&=(H\otimes 1)\frac{1}{\sqrt{d}}\sum\limits_{k=0}^{d-1}\exp\left(\frac{2\pi i k (x-y)}{d}\right)|ky\rangle\\
&= |x-y,y\rangle.
\end{aligned}
\end{equation}
Together with arbitrary qudit rotations and measurements in the qudit $X$ and $Z$ basis, 
this suffices to implement a deterministic Bell state analyzer for qudits \cite{Dusek}.
\section{Discussion and Conclusions}
\label{sec: conclusions}
We introduced a hybrid quantum repeater protocol for the distribution of arbitrary finite-dimensional bipartite
entangled states over large distances with a specific focus on qutrit entanglement. A generalization of the dispersive light-matter interaction
from the qubit to the general qudit case lies at the heart of our protocol and can be expressed by higher 
spin operators.
The distribution of matter-matter entanglement between neighboring repeater stations is mediated via coherent states interacting dispersively and subsequently with the matter systems. We investigated both USD 
and homodyne detection of the light mode and compared the rates and final fidelities. By exploiting purification on the elementary segments, sufficiently
high initial fidelities can be achieved to cover distances up to 1280 km with final fidelities close to unity. With three rounds of 
entanglement purification directly after the initial entanglement distributions, rates $\sim$ 100 Hz are, in principle, possible.\\
Since our scheme assumes perfect matter systems (with perfect coherence properties for arbitrarily long times) and operations on them, future research may aim at investigating
different physical platforms and decoherence models for the matter systems. Like for the qubit case \cite{hybridencoding}, 
quantum error correction codes could be employed on the matter systems turning the scheme to a genuine second generation quantum repeater scheme and thus
preserving the communication rates obtained here under idealizing assumptions.
\section{Acknowledgement}
\noindent We acknowledge support from Q.com (BMBF).\\

\bibliographystyle{apsrev4-1}
\bibliography{hybridqudit.bib}
\newpage
\appendix
\onecolumngrid
\section{Rate analysis for qutrit hybrid quantum repeater}
\label{sec: rate}
In this appendix, we show several tables summarizing the results on the rates
and fidelities for our qutrit quantum repeater scheme ($d=3$), as described
in Sec. \ref{sec: rate2}. We consider various total distances 
up to 1280 km, two possible elementary distances ($L_{0}=5, 10$ km),
between zero and three rounds of entanglement purification directly after the initial entanglement distribution, and
the two possible detection schemes (homodyne, USD).
\begin{table}[htp]
\begin{ruledtabular}
  \begin{tabular}{c|c|c|c|c|c|}
  &rounds of purification&no&one&two&three\\
  &initial fidelity&0.75&	0.94393&0.997854&0.999996\\
  &effective probability&0.64&0.302641&0.19154&0.1318\\
  \hline
  rate [Hz]&10 km&10175&4290&2647&900\\
  &20 km&7936&3185&1942&656\\
  &40 km&6366&2488&1506&507\\
  &80 km&5285&2024&1220&409\\
  &160 km&4501&1701&1021&342\\
  &320 km&3914&1464&877&294\\
  &640 km&3461&1284&768&257\\
  \hline
  fidelity&10 km&0.56&0.891&0.9957&0.99999265\\
  &20 km&0.315&	0.793&0.9914&0.999998531\\
  &40 km&0.09&0.63&0.983&0.99997061\\
  &80 km&0&0.397&0.966&0.99994123\\
  &160 km&0&0.158&0.934&0.99988246\\
  &320 km&0&0.02&0.872&0.99976494\\
  &640 km&0&0&0.761&0.99952994
  \end{tabular}
  \caption{$L_{0}$=5 km (USD), $\alpha=1.2$, $L\leq$ 640 km}
  \label{table:5USD}
\end{ruledtabular}
\end{table}

\begin{table}[htp]
\begin{ruledtabular}
  \begin{tabular}{|c|c|c|c|c|c|}
  &rounds of purification&no&one&two&three\\
  \hline
  &initial fidelity&0.652&0.87&	0.987&0.999\\
  \hline
  &effective probability&0.414&	0.147&0.078&0.05\\
  \hline
  rate [Hz]&20 km&3020&1010&524&343\\
           &40 km&2271&738&380&248\\
           &80 km&1788&570&293&191\\
           &160 km&1463&461&236&156\\
           &320 km&1234&385&197&128\\
           &640 km&1065&331&169&110\\
           &1280 km&936&289&147&96\\
  \hline
  fidelity&20 km&0.420&	0.76&0.974&0.999\\
  &40 km&0.18&0.57&0.95&0.999\\
  &80 km&0.03&0.33&0.9&	0,999\\
  &160 km&0.001&0.1&0.814&0.998\\
  &320 km&0&	0.01&	0.66&	0.996\\
  &640 km&0&0&0.436&0.992\\
  &1280 km&0&0&0.19&0.984
  \end{tabular}
  \caption{$L_{0}$=10 km (USD), $\alpha=1.2$, $L\leq$ 1280 km}
  \label{table:10USD}
\end{ruledtabular}
\end{table}

\begin{table}[htp]
\begin{ruledtabular}
  \begin{tabular}{|c|c|c|c|c|c|}
  &rounds of purification&no&one&two&three\\
  \hline
  &initial fidelity&0.73&0.93&0.997&0.999997\\
  \hline
  &effective probability&0.38&0.15&0.09&0.0619534\\
  \hline
  rate [Hz]&10 km&5496&2056&1219&835\\
           &20 km&4117&1502&885&605\\
           &40 km&3233&1161&682&465\\
           &80 km&2641&939&550&375\\
           &160 km&2225&785&459&313\\
           &320 km&1919&674&394&267\\
           &640 km&1686&589&344&234\\
  \hline
  fidelity&10 km&0.53&0.86&0.995&0.999994\\
  &20 km&0.28&0.75&0.990&0.999987\\
  &40 km&0.08&0.56&0.980&0.999975\\
  &80 km&0.01&0.31&0.961&0.99995\\
  &160 km&0.00&0.10&0.923&0.9999\\
  &320 km&0.00&0.01&0.852&0.9998\\
  &640 km&0.00&0.00&0.726&0.9996
  \end{tabular}
  \caption{$L_{0}$=5 km (homodyne), $\alpha\approx 1$, $L\leq$ 640 km}
  \label{table:5x}
\end{ruledtabular}
\end{table}

\begin{table}[htp]
\begin{ruledtabular}
  \begin{tabular}{|c|c|c|c|c|c|}
  &rounds of purification&no&one&two&three\\
  \hline
  &initial fidelity&0.6&0.81&0.974&0.9996\\
  \hline
  &effective probability&0.39&0.12&0.057&0.037\\
  \hline
  rate [Hz]&20 km&2828&817&384&246\\
           &40 km&2121&595&278&178\\
           &80 km&1667&460&214&137\\
           &160 km&1362&371&172&110\\
           &320 km&1148&310&144&92\\
           &640 km&990&266&123&79\\
           &1280 km&870&233&107&69\\
  \hline
  fidelity&20 km&0.360&	0.656&0.949&0.999\\
  &40 km&0.130&	0.430&0.900&0.999\\
  &80 km&0.017&0.185&0.810&0.997\\
  &160 km&0.000&0.034&0.656&0.994\\
  &320 km&0.000&0.001&0.430&0.989\\
  &640 km&0&0&0.184&0.978\\
  &1280 km&0&0&0.03&0.957
  \end{tabular}
  \caption{$L_{0}$=10 km (homodyne), $\alpha\approx 1$, $L\leq$ 1280 km}
  \label{table:10x}
\end{ruledtabular}
\end{table}

\begin{table}[htp]
\begin{ruledtabular}
  \begin{tabular}{|c|c|c|c|c|}
  &rounds of purification&no&one&two\\
  \hline
  &initial fidelity&0.861808&0.986275&0.999876\\
  \hline
  &effective probability&0.0137597&0.0069238&0.0044958\\
  \hline
  rate [Hz]&40 km&92&46&30\\
  &80 km&33&17&11\\
  &160 km&26&13&9\\
  &320 km&21&11&7\\
  &640 km&17&9&6\\
  &1280 km&15&8&5\\
  \hline
  fidelity&20 km&0.360&	0.656&0.949\\
  &40 km&0.130&	0.973&0.9997\\
  &80 km&0.017&0.946&0.9995\\
  &160 km&0.000&0.895&0.9990\\
  &320 km&0.09&0.802&0.9980\\
  &640 km&0&0&0.9960\\
  &1280 km&0&0&0.9921
  \end{tabular}
  \caption{$L_{0}$=20 km (USD), $\alpha=0.5$, $L\leq$ 1280 km}
  \label{table:20x}
\end{ruledtabular}
\end{table}

\clearpage
\newpage
\section{Ququart hybrid repeater}
\label{sec:ququart}
The dispersive interaction acting on a ququart-light system is defined 
by the unitary transformation
 \begin{equation}
 \begin{aligned}
 U_{4}(\theta)&\left[\frac{1}{2}(|0\rangle+|1\rangle+|2\rangle+|3\rangle)|\alpha\rangle\right]
 =\frac{1}{2}(|0\rangle|\alpha\rangle+|1\rangle|\alpha e^{i\theta}\rangle
 +|2\rangle|\alpha e^{2i\theta}\rangle+|3\rangle|\alpha e^{3i\theta}\rangle),
 \end{aligned}
 \end{equation}
 which is induced by the Hamiltonian $H_{int}^{(4)}=\hbar g S^{(4)}_{z} a^{\dagger}a$
 with $S^{(4)}_{z}|k\rangle=\left(\frac{2k-3}{2}\right)|k\rangle$ for $k\in \{0,1,2,3\}$.
 Thus, the ququart (4-level system) may be represented by a spin-$\frac{3}{2}$ particle.
 The case of a strong interaction is obtained by choosing $\theta=\frac{\pi}{2}$.\\
 As before, the first step in the protocol is the generation of an entangled ququart-light state
 via the strong dispersive interaction, i.e.,
 \begin{equation}
  \frac{1}{2}(|0\rangle|\alpha\rangle+|1\rangle|i\alpha\rangle
 +|2\rangle|-\alpha\rangle+|3\rangle|-i\alpha \rangle),
\end{equation}
of which the light part is then sent through the optical channel over a distance $L_{0}$,
suffering from loss.\\
The output density matrix is again determined by mixing the light mode with
an ancilla vacuum state and tracing out the light mode. It is again useful to transform
the coherent states of the light field into an orthogonal basis.
The adapted orthogonal basis in this case reads
 \begin{equation}
 \label{eq: ququartbasis}
 \begin{aligned}
 |u\rangle&=\frac{1}{\sqrt{N_{u}(\alpha)}}(|\alpha\rangle+|-\alpha\rangle+|i\alpha\rangle+|-i\alpha\rangle),\\
 |v\rangle&=\frac{1}{\sqrt{N_{v}(\alpha)}}(|\alpha\rangle+i|-\alpha\rangle-|i\alpha\rangle-i|-i\alpha\rangle),\\
 |w\rangle&=\frac{1}{\sqrt{N_{w}(\alpha)}}(|\alpha\rangle-|-\alpha\rangle+|i\alpha\rangle-|-i\alpha\rangle),\\
 |z\rangle&=\frac{1}{\sqrt{N_{z}(\alpha)}}(|\alpha\rangle-i|-\alpha\rangle-|i\alpha\rangle+i|-i\alpha\rangle),\\
\end{aligned}
\end{equation}
with normalization constants $N_{u}(\alpha), N_{v}(\alpha),N_{w}(\alpha),$ and $N_{z}(\alpha)$. We can therefore write
\begin{equation}
\begin{aligned}
 |\alpha\rangle&=\frac{1}{4}(\sqrt{N_{u}(\alpha)}|u\rangle+\sqrt{N_{v}(\alpha)}|v\rangle+\sqrt{N_{w}(\alpha)}|w\rangle+\sqrt{N_{z}(\alpha)}|z\rangle),\\
 |-\alpha\rangle&=\frac{1}{4}(\sqrt{N_{u}(\alpha)}|u\rangle-i\sqrt{N_{v}(\alpha)}|v\rangle-\sqrt{N_{w}(\alpha)}|w\rangle+i\sqrt{N_{z}(\alpha)}|z\rangle),\\
 |i\alpha\rangle&=\frac{1}{4}(\sqrt{N_{u}(\alpha)}|u\rangle-\sqrt{N_{v}(\alpha)}|v\rangle+\sqrt{N_{w}(\alpha)}|w\rangle-\sqrt{N_{z}(\alpha)}|z\rangle),\\
 |-i\alpha\rangle&=\frac{1}{4}(\sqrt{N_{u}(\alpha)}|u\rangle+i\sqrt{N_{v}(\alpha)}|v\rangle-\sqrt{N_{w}(\alpha)}|w\rangle-i\sqrt{N_{z}(\alpha)}|z\rangle).
 \end{aligned}
 \end{equation}
 
The resulting output density matrix, 
\begin{equation} 
\label{eq: ququartout}
\begin{aligned}
  \rho_{out}&=\frac{N_{u}(\sqrt{1-\gamma}\alpha)}{16} 
  \times\left[\frac{1}{2}(|0\rangle|\sqrt{\gamma}\alpha\rangle+|1\rangle|-\sqrt{\gamma}\alpha\rangle+|2\rangle|i\sqrt{\gamma}\alpha\rangle
 +|3\rangle|-i\sqrt{\gamma}\alpha\rangle)\right]\times H.c.\\
 &+\frac{N_{v}(\sqrt{1-\gamma}\alpha)}{16}\times\left[\frac{1}{2}(|0\rangle|\sqrt{\gamma}\alpha\rangle-i|1\rangle|-\sqrt{\gamma}\alpha\rangle-|2\rangle|i\sqrt{\gamma}\alpha\rangle+i|3\rangle|-i\sqrt{\gamma}\alpha\rangle)\right]\times H.c.\\
 &+\frac{N_{w}(\sqrt{1-\gamma}\alpha)}{16}\times\left[\frac{1}{2}(|0\rangle|\sqrt{\gamma}\alpha\rangle-|1\rangle|-\sqrt{\gamma}\alpha\rangle+|2\rangle|i\sqrt{\gamma}\alpha\rangle-|3\rangle|-i\sqrt{\gamma}\alpha\rangle)\right]\times H.c.\\
 &+\frac{N_{z}(\sqrt{1-\gamma}\alpha)}{16}\times\left[\frac{1}{2}(|0\rangle|\sqrt{\gamma}\alpha\rangle+i|1\rangle|-\sqrt{\gamma}\alpha\rangle-|2\rangle|i\sqrt{\gamma}\alpha\rangle-i|3\rangle|-i\sqrt{\gamma}\alpha\rangle)\right]\times H.c.,
 \end{aligned}
 \end{equation}
 is now a four-component mixture. This entangled ququart-light state can be further
 simplified by switching to the orthogonal basis (Eq. \eqref{eq: ququartbasis})
 for the light mode and to the \textit{X}-Basis
 \begin{equation}
  \begin{aligned}
   |\widetilde{0}\rangle&=\frac{1}{2}(|0\rangle+|1\rangle+|2\rangle+|3\rangle),\\
   |\widetilde{1}\rangle&=\frac{1}{2}(|0\rangle+i|1\rangle-|2\rangle-i|3\rangle),\\
   |\widetilde{2}\rangle&=\frac{1}{2}(|0\rangle-|1\rangle+|2\rangle-|3\rangle),\\
   |\widetilde{3}\rangle&=\frac{1}{2}(|0\rangle-i|1\rangle-|2\rangle+i|3\rangle),
  \end{aligned}
\end{equation}
for the matter system. Using these bases, Eq. \eqref{eq: ququartout}
can be rewritten as
\begin{equation}
\begin{aligned}
 \rho_{out}=&\frac{N_{u}(\sqrt{1-\gamma}\alpha)}{16}
 \times\left[\frac{1}{4}\left(\sqrt{N_{u}(\sqrt{\gamma}\alpha)}|\widetilde{0}\rangle|\widetilde{u}\rangle
 +\sqrt{N_{v}(\sqrt{\gamma}\alpha)}|\widetilde{1}\rangle|\widetilde{v}\rangle\right.
  +\sqrt{N_{w}(\sqrt{\gamma}\alpha)}|\widetilde{2}\rangle|\widetilde{w}\rangle
   +\sqrt{N_{z}(\sqrt{\gamma}\alpha)}|\widetilde{3}\rangle|\widetilde{z}\rangle
 \left.\right)\right]\times H.c.\\
 +&\frac{N_{v}(\sqrt{1-\gamma}\alpha)}{16}
 \times\left[\frac{1}{4}\left(\sqrt{N_{u}(\sqrt{\gamma}\alpha)}|\widetilde{3}\rangle|\widetilde{u}\rangle\right.
 +\sqrt{N_{v}(\sqrt{\gamma}\alpha)}|\widetilde{2}\rangle|\widetilde{v}\rangle
  +\sqrt{N_{w}(\sqrt{\gamma}\alpha)}|\widetilde{1}\rangle|\widetilde{w}\rangle
   +\sqrt{N_{z}(\sqrt{\gamma}\alpha)}|\widetilde{0}\rangle|\widetilde{z}\rangle
 \left.\right)\right]\times H.c.\\
  +&\frac{N_{w}(\sqrt{1-\gamma}\alpha)}{16}
 \times\left[\frac{1}{4}\left(\sqrt{N_{u}(\sqrt{\gamma}\alpha)}|\widetilde{2}\rangle|\widetilde{u}\rangle\right.
 +\sqrt{N_{v}(\sqrt{\gamma}\alpha)}|\widetilde{1}\rangle|\widetilde{v}\rangle
  +\sqrt{N_{w}(\sqrt{\gamma}\alpha)}|\widetilde{0}\rangle|\widetilde{w}\rangle
   +\sqrt{N_{z}(\sqrt{\gamma}\alpha)}|\widetilde{3}\rangle|\widetilde{z}\rangle
 \left.\right)\right]\times H.c.\\
  +&\frac{N_{z}(\sqrt{1-\gamma}\alpha)}{16}
 \times\left[\frac{1}{4}\left(\sqrt{N_{u}(\sqrt{\gamma}\alpha)}|\widetilde{1}\rangle|\widetilde{u}\rangle\right.
 +\sqrt{N_{v}(\sqrt{\gamma}\alpha)}|\widetilde{0}\rangle|\widetilde{v}\rangle
  +\sqrt{N_{w}(\sqrt{\gamma}\alpha)}|\widetilde{3}\rangle|\widetilde{w}\rangle
   +\sqrt{N_{z}(\sqrt{\gamma}\alpha)}|\widetilde{2}\rangle|\widetilde{z}\rangle
\left. \right)\right]\times H.c.
 \end{aligned}
 \end{equation}
where $\sim$ again indicates basis vectors with damped amplitudes for the light-mode states.\\
The light mode of the state in Eq. \eqref{eq: ququartout} finally interacts 
with a second matter system via the inverse dispersive interaction with $\theta=-\frac{\pi}{2}$.
The resulting state reads
 \begin{equation}
 \begin{aligned}
  \rho&=\frac{N_{u}(\sqrt{1-\gamma}\alpha)}{16}|D_{0}\rangle\langle D_{0}|
 +\frac{N_{v}(\sqrt{1-\gamma}\alpha)}{16} |D_{1}\rangle\langle D_{1}|\\
 &+\frac{N_{w}(\sqrt{1-\gamma}\alpha)}{16} |D_{2}\rangle\langle D_{2}|
 +\frac{N_{z}(\sqrt{1-\gamma}\alpha)}{16} |D_{3}\rangle\langle D_{3}|,\\
 \end{aligned}
 \end{equation}
 with the components
 \begin{equation}
 \begin{aligned}
 |D_{0}\rangle&=\frac{1}{2}\left(\frac{1}{2}(|00\rangle+|11\rangle+|22\rangle+|33\rangle)|\sqrt{\gamma}\alpha\rangle\right.\\
 &+\frac{1}{2}(|01\rangle+|10\rangle+|23\rangle+|32\rangle)|-\sqrt{\gamma}\alpha\rangle\\
 &+\frac{1}{2}(|03\rangle+|12\rangle+|20\rangle+|31\rangle)|i\sqrt{\gamma}\alpha\rangle\\
 &\left.+\frac{1}{2}(|02\rangle+|13\rangle+|21\rangle+|30\rangle)|-i\sqrt{\gamma}\alpha\rangle\right),
\end{aligned}
\end{equation}

\begin{equation}
\begin{aligned}
 |D_{1}\rangle&=\frac{1}{2}\left(\frac{1}{2}(|00\rangle-i|11\rangle-|22\rangle+i|33\rangle)\right.|\sqrt{\gamma}\alpha\rangle\\
 &+\frac{1}{2}(|01\rangle-i|10\rangle-|23\rangle+i|32\rangle)|-\sqrt{\gamma}\alpha\rangle\\
 &+\frac{1}{2}(|03\rangle-i|12\rangle-|20\rangle+i|31\rangle)|i\sqrt{\gamma}\alpha\rangle\\
 &+\left.\frac{1}{2}(|02\rangle-i|13\rangle-|21\rangle+i|30\rangle)|-i\sqrt{\gamma}\alpha\rangle\right),
\end{aligned}
\end{equation}
\begin{equation}
\begin{aligned}
 |D_{2}\rangle&=\frac{1}{2}\left(\frac{1}{2}(|00\rangle-i|11\rangle-|22\rangle+i|33\rangle)|\sqrt{\gamma}\alpha\rangle\right.\\
 &+\frac{1}{2}(|01\rangle-i|10\rangle-|23\rangle+i|32\rangle)|-\sqrt{\gamma}\alpha\rangle\\
 &+\frac{1}{2}(|03\rangle-i|12\rangle-|20\rangle+i|31\rangle)|i\sqrt{\gamma}\alpha\rangle\\
 &\left.+\frac{1}{2}(|02\rangle-i|13\rangle-|21\rangle+i|30\rangle)|-i\sqrt{\gamma}\alpha\rangle\right),
\end{aligned}
\end{equation}

\begin{equation}
\begin{aligned}
 |D_{3}\rangle&=\frac{1}{2}\left(\frac{1}{2}(|00\rangle+i|11\rangle-|22\rangle-i|33\rangle)|\sqrt{\gamma}\alpha\rangle\right.\\
 &+\frac{1}{2}(|01\rangle+i|10\rangle-|23\rangle-i|32\rangle)|-\sqrt{\gamma}\alpha\rangle\\
 &+\frac{1}{2}(|03\rangle+i|12\rangle-|20\rangle-i|31\rangle)|i\sqrt{\gamma}\alpha\rangle\\
 &\left.+\frac{1}{2}(|02\rangle+i|13\rangle-|21\rangle-i|30\rangle)|-i\sqrt{\gamma}\alpha\rangle\right).
\end{aligned}
\end{equation}
The remaining task is then to project onto the coherent states
$|\sqrt{\gamma}\alpha\rangle,|-\sqrt{\gamma}\alpha\rangle, |i\sqrt{\gamma}\alpha\rangle$ and $|-i\sqrt{\gamma}\alpha\rangle$
to establish a maximally entangled state in each of the components.
Due to the special structure of the coherent states under consideration, homodyne detection
in the ququart case is more problematic than in the qutrit case.\\
The states $|\pm \sqrt{\gamma}\alpha\rangle$ have Gaussian position distribution around $\pm \sqrt{\gamma}\alpha$,
whereas $|\pm i\sqrt{\gamma}\alpha\rangle$ are both distributed around zero and therefore cannot be distinguished by an $x$-measurement.
The same is true for a $p$-measurement, where $|\pm \sqrt{\gamma}\alpha\rangle$ have now both average zero and
$|\pm i\sqrt{\gamma}\alpha\rangle$ have means $\sqrt{\gamma}\alpha$ and $-\sqrt{\gamma}\alpha$, respectively. Therefore,
deterministic entanglement generation is not possible and the corresponding terms 
in the superposition have to be discarded.\\
If we choose the $x$-measurement, the selection windows are then the same as in the qubit case:
$w_{0}=[\sqrt{\gamma}\alpha-\Delta,\infty]$ with $\Delta>0$ 
corresponds to a projection onto $\frac{1}{2}(|00\rangle+|11\rangle+|22\rangle+|33\rangle)$,
whereas a measurement result in $w_{1}=[-\infty, -\sqrt{\gamma}\alpha+\Delta]$
leads to $\frac{1}{2}(|01\rangle+|10\rangle+|23\rangle+|32\rangle)$. In both cases, of course,
an error due to the non-orthogonality of the coherent states has to be taken into account.\\
The probability for optimally distinguishing the four coherent states via USD
as well as entanglement purification and swapping are addressed in Sec. \ref{sec: qudit} in a 
as a special case of the general qudit.

\end{document}